\documentclass[12pt]{article}
\usepackage{cite}
\usepackage{graphicx}
\usepackage{amssymb,amsfonts}
\usepackage{hangcaption}

\renewcommand{\theequation}
{\arabic{section}.\arabic{equation}}

\makeatletter
\@addtoreset{equation}{section}
\@addtoreset{footnote}{section}
\makeatother

\makeatletter
\renewcommand\appendix{\par
  \setcounter{section}{0}%
  \setcounter{subsection}{0}%
  \gdef\thesection{Appendix \@Alph\c@section }
  \renewcommand{\theequation}
  {\Alph{section}.\arabic{equation}}
}
\makeatother

\makeatletter
\def\eqnarray{ \stepcounter{equation} \let\@currentlabel=\theequation
 \global\@eqnswtrue
 \global\@eqcnt\z@
 \tabskip\@centering
 \let\\=\@eqncr
 $$\halign to \displaywidth\bgroup\@eqnsel\hskip\@centering
 $\displaystyle\tabskip\z@{##}$&\global\@eqcnt\@ne
 \hfil$\displaystyle{{}##{}}$\hfil
 &\global\@eqcnt\tw@$\displaystyle\tabskip\z@{##}$\hfil
 \tabskip\@centering&\llap{##}\tabskip\z@\cr}
\makeatother

\makeatletter
\def\@arrayacol{\edef\@preamble{\@preamble \hskip .5\arraycolsep}}
\def\array{\let\@acol\@arrayacol \let\@classz\@arrayclassz
\let\@classiv\@arrayclassiv \let\\\@arraycr\def\@halignto{}\@tabarray}
\makeatother

\renewcommand{\arraystretch}{1.6}

\makeatletter
\newcounter{subeqncnt}
\def\thesubeqncnt{\alph{subeqncnt}}
\def\subequations{\begingroup%
   \stepcounter{equation}\edef\@tempa{\theequation}%
   \let\c@equation\c@subeqncnt\c@subeqncnt\z@
   \edef\theequation{\@tempa\noexpand\thesubeqncnt}}

\makeatother

\newcommand{\captionfonts}{\small} 
\makeatletter 
\long\def\@makecaption#1#2{%
\vskip\abovecaptionskip
\sbox\@tempboxa{{\captionfonts #1: #2}}%
\ifdim \wd\@tempboxa >\hsize
{\captionfonts #1: #2\par}
\else
\hbox to\hsize{\hfil\box\@tempboxa\hfil}%
\fi
\vskip\belowcaptionskip}
\makeatother 

\newcommand{\del}{\partial}

\newcommand{\tr}{{\rm Tr}\,}
\newcommand{\dd}{{\rm d}}

\newcommand{\tC}{{\widetilde C}}
\newcommand{\tD}{{\widetilde D}}

\begin{document}
\setlength{\baselineskip}{7mm}
\begin{titlepage}
\begin{flushright}
{\tt APCTP-Pre2009-001} \\
{\tt arXiv:0901.0610[hep-th]}\\
January 2009
\end{flushright}

\vspace{1cm}

\begin{center}

{\Large   
Sound Modes in Holographic Hydrodynamics \\
for Charged AdS Black Hole}

\vspace{1cm}

{\sc{Yoshinori Matsuo}}$^*$,
{\sc{Sang-Jin Sin}}$^\dagger{}^*$, \\
{\sc{Shingo Takeuchi}}$^*$, 
{\sc{Takuya Tsukioka}}$^*$
and 
{\sc{Chul-Moon Yoo}}$^*$

\vspace*{3mm}

$*$
{\it{Asia Pacific Center for Theoretical Physics},} \\
{\it{Pohang, Gyeongbuk 790-784, Korea}} \\
{\sf{ymatsuo, shingo, tsukioka, c\_m\_yoo@apctp.org}}
\\
$\dagger$
{\it{Department of Physics,}}
{\it{Hanyang University,}} 
{\it{Seoul 133-791, Korea}} \\
{\sf{sjsin@hanyang.ac.kr}}
\end{center}

\vspace{1cm}

\begin{abstract}
In the previous paper
we studied the transport coefficients of Quark-Gluon-Plasma 
in finite temperature and finite density in vector and tensor modes. 
In this paper, we extend it to the scalar modes. 
We work out the decoupling problem and hydrodynamic analysis 
for the sound mode in charged AdS black hole 
and calculate the sound velocity, the charge susceptibility 
and the electrical conductivity. 
We find that Einstein relation among the conductivity, 
the diffusion constant and the susceptibility holds exactly. 
\end{abstract} 

\end{titlepage}

\section{Introduction}

The discovery of low viscosity in the theory with gravity dual~\cite{pss0} 
and its possible relation to the RHIC (Relativistic Heavy Ion Collider) 
experiment induced a great deal of efforts to establish 
the relevant calculational scheme  that may be provided by  
AdS/CFT correspondence~\cite{ads/cft, gkp, w}. 
An attempt has been made to map the entire process 
of RHIC experiment in terms of the gravity dual~\cite{ssz}.  
The way to include a chemical potential in the theory was 
figured out in the context of probe brane embedding \cite{ksz,ht,nssy1,kmmmt,nssy2,bergman,ubc,n}.  
Phases of these theories were discussed  and 
new phases were reported where instability due to the strong 
attraction is a feature~\cite{nssy1,kmmmt,nssy2}. 

In spite of the difference between QCD and ${\cal N}=4$ SYM, 
it is expected that some of the properties are shared 
by the two theories based on the universality of low energy physics.
In this respect, the hydrodynamic limit is particularly interesting 
since such limit can be shared by many theories. 
The calculation scheme for transport coefficients 
is to use Kubo formula, which gives a relation to the low energy 
limit of Wightman Green functions.
In AdS/CFT correspondence, one calculate the retarded 
Green function which is related  to 
the Wightman function by fluctuation-dissipation theorem.
Such scheme has been developed 
in a series of papers~\cite{ss,pss,pss2,hs,ks}.
 
For the hydrodynamic analysis, one may need to have 
master equations for the 
decoupled modes in vector and scalar at hands. 
Although the analysis for the decoupling problem were analyzed 
in~\cite{ki}, it was based on $SO(3)$ decomposition 
while more useful work for hydrodynamics  should be based on 
$SO(2)$ decomposition, where longitudinal direction of the 
spatial direction is distinguished. 
For this purpose, Kovtun and Starinets worked out the 
decoupling problem based on $SO(2)$ for the AdS black hole 
case~\cite{ks} before doing the hydrodynamic analysis. 
For the charged cases, there are extra difficulties: 
vector modes of gravity and those of the gauge fields couple. 
Furthermore there are extra couplings in scalar modes 
which are not present in the chargeless cases. 

In the previous paper~\cite{gmsst}, some of us 
extended this work to charged case using the 
Reissner-Nordstr{\"o}m-Anti-deSitter (RN-AdS) black hole, 
which corresponds to the diagonal $(1,1,1)$  R-charged STU 
black hole\footnote{
In fact much works had been done  for charged case 
by various groups~\cite{mas,ss2,mno, saremi, bbn}.
In~\cite{mas,ss2},   thermodynamics for STU black 
hole~\cite{cvetic,cvetic2} and the hydrodynamic calculations 
 for the $(1,0,0)$ charge were performed. 
In~\cite{mno, saremi}, charged AdS$_5$ and AdS$_4$ black hole 
backgrounds were considered, respectively, and it was shown 
numerically that the ratio $(\eta/s)$ 
was $1/(4\pi)$ with very good accuracy. 
Later, it was also proven that the ratio might be universal in more general 
setup~\cite{bbn}. 
}.
However, analysis for the scalar mode of charged case was 
not done due to difficulties caused by extra mixing between 
various scalar modes in charged AdS black hole. 
In this paper, we work out the  decoupling problem and hydrodynamics 
for the sound (scalar) mode in such case. 
Green functions are explicitly obtained.  
Our results show that the behavior of the transport coefficients 
in RN-black hole are very different from those in the 
(1,0,0) charged black hole: the formers are much more smoother 
than the latters. 
We find that Einstein relation among the conductivity, the diffusion 
constant and the susceptibility holds exactly. 
  
This paper is organized as follows: 
In Section 2, we introduce RN-AdS black hole and review 
correlation function calculation at finite temperature in 
AdS/CFT correspondence. 
In Section 3, a formulation on the metric and the gauge perturbations 
in RN-AdS background is considered. 
We then solve linearized perturbative equations of motion in 
hydrodynamic regime and 
obtain retarded Green functions in Section 4. 
We also observe the transport coefficients including 
the speed of sound, the diffusion constant for $U(1)$ charge and 
the electrical conductivity in this section. 
Conclusions and discussions are given in the final section. 
Three appendices are provided. 
In Appendix A, we summarize the results in the vector and the 
tensor type perturbations in our previous work~\cite{gmsst}.
The details of calculations to solve equations of motion are given
in Appendix B and C.     

\section{Basic Setup}\label{sec:Setup}

\subsection{Minkowskian correlators in AdS/CFT correspondence}

Before introducing RN-AdS black hole, 
we briefly summarize to obtain Minkowskian correlators 
in AdS/CFT correspondence. 
We follow the prescription proposed in~\cite{ss}.
We work on the five-dimensional background,  
\begin{equation}
 \dd s^2 = g_{\mu\nu}\dd x^\mu\dd x^\nu + g_{uu}(\dd u)^2,  
\end{equation}
where $x^\mu$ and $u$ are the four-dimensional and the radial coordinates, 
respectively. 
We refer the boundary as $u=0$ and the horizon as $u=1$.  
A solution of the equation of motion may be given,  
\begin{equation}
\phi(u,x) = 
\!\int\!\frac{\dd^4 k}{(2\pi)^4}\ \mbox{e}^{ikx}f_k(u)\phi^0(k), 
\end{equation}
where $f_k(u)$ is normalized such that $f_k(0)=1$ at the boundary.  
An on-shell action might be reduced to surface terms 
by using the equation of motion, 
\begin{equation}
S[\phi^0]
=\!\int\!\frac{\dd^4 k}{(2\pi)^4}
\phi^0(-k){\cal G}(k, u)\phi^0(k)
\bigg|_{u=0}^{u=1} . 
\label{on_shell_action}
\end{equation}
Here, the function $\mathcal G(k,u)$ 
can be written in terms of $f_{\pm k}(u)$ and $\partial_u f_{\pm k}(u)$.  
Accommodating 
Gubser-Klebanov-Polyakov/Witten relation~\cite{gkp,w} 
to Minkowski spacetime, 
Son and Starinets proposed 
the formula to get retarded Green functions,  
\begin{equation}
G^{\rm R}(k)
=
2{\cal G}(k, u)
\bigg|_{u=0}, 
\label{green_function}
\end{equation}
where the incoming boundary condition at the horizon is imposed.  
In general, there are several fields in the model. 
We write the Green function as $G_{ij}(k)$, 
where indices $i$ and $j$ distinguish corresponding fields. 

In this paper, we work in RN-AdS background and 
consider its perturbations so that 
essential ingredients are perturbed metric field and $U(1)$ 
gauge field.  
Here we define the precise form of the retarded Green functions 
which we discuss later:   
%
\renewcommand{\arraystretch}{2.0}
%
\begin{eqnarray}
\begin{array}{rcl}
G_{\mu\nu \ \rho\sigma}(k)
&=&
\displaystyle
-i\!\int\!\dd^4x
\ \mbox{e}^{-ikx}\theta(t)
\langle{[}T_{\mu\nu}(x), \ T_{\rho\sigma}(0){]}\rangle, 
\\
G_{\mu\nu \ \rho}(k)
&=&
\displaystyle
-i\!\int\!\dd^4x
\ \mbox{e}^{-ikx}\theta(t)
\langle{[}T_{\mu\nu}(x), \ J_\rho(0){]}\rangle, 
\\
G_{\mu \ \nu}(k)
&=&
\displaystyle
-i\!\int\!\dd^4x
\ \mbox{e}^{-ikx}\theta(t)
\langle{[}J_\mu(x), \ J_\nu(0){]}\rangle, 
\end{array}
\label{diff_green_function}
\end{eqnarray}
%
\renewcommand{\arraystretch}{1.6}
%

\vspace*{-3mm}
\noindent
where the operators $T_{\mu\nu}(x)$ and $J_\mu(x)$ are energy-momentum 
tensor and $U(1)$ current which couple to the metric and the gauge field,  
respectively.  

\subsection{Reissner-Nordstr{\"o}m-AdS background}

The charge in RN-AdS black hole is usually regarded 
as $R$-charge of SUSY~\cite{myers}. 
We here consider an another interpretation in the following way:   
One can introduce quarks and mesons by considering 
the bulk-filling branes in AdS$_5$ space. 
The overall $U(1)$ of the flavor branes is identified 
as the baryon charge. 
The $U(1)$ charge in this model~\cite{s} minimally couples 
to the bulk gravity since the bulk and the world volume of brane  
are identified. Then, the baryon charge and the $R$-charge have the same description in terms of the $U(1)$ gauge field living in the AdS${}_5$ space.   
A charged black hole (RN-AdS black hole) is then induced by its back 
reaction. 
Therefore the $U(1)$ charge in RN-AdS can be identical to the baryon 
charge.
As a result, we can interpret  our  result as a calculation of
the transport coefficients in the presence of the baryon density. 

The effective action of this gauge field is given 
the quadratic piece of Dirac-Born-Infeld action\footnote{
The indices $m$ and $n$ run through five-dimensional spacetime
while $\mu$ and $\nu$ would be reserved for four-dimensional Minkowski
spacetime. Their spatial coordinates are labeled by $i$ and $j$.
}
\begin{equation}
 S_{\mathrm{DBI}} 
= 
-\frac{1}{4e^2}
\!\int\!\dd^5x\sqrt{-g}\ 
\tr\bigl({\cal F}_{mn}{\cal F}^{mn}\bigl) ,  
\end{equation}
where the gauge coupling constant $e$ is given by \cite{s} 
\begin{equation}
 \frac{l}{e^2} = \frac{N_cN_f}{4\pi^2} , 
\label{le2}
\end{equation}
with $l$ the radius of the AdS space.  
We pick up an overall $U(1)$ part of this gauge field in order to 
consider the baryon current at the boundary.  
Together with the gravitation part, 
we arrive at the following action which is our starting point:  
\begin{equation}
S_0
=\frac{1}{2\kappa^2}\!\int\!\dd^5x\sqrt{-g}
\Big(
R-2\Lambda
\Big)
-
\frac{1}{4e^2}
\!\int\!\dd^5x\sqrt{-g}
{\cal F}_{mn}{\cal F}^{mn},
\label{action_bh}
\end{equation}
where we denote the gravitation constant and the cosmological constant 
as $\kappa^2=8\pi G_5$ and $\Lambda$, respectively.  
The $U(1)$ gauge field strength is given by 
${\cal F}_{mn}(x)=\del_m{\cal A}_n(x)-\del_n{\cal A}_m(x)$. 
The gravitation constant is related to the gauge theory quantities by 
\begin{equation}
 \frac{l^3}{\kappa^2} = \frac{N_c^2}{4\pi^2} . 
\end{equation}
Suppose we have baryon charge $Q$. 
This should be identified to the source of $U(1)$ charge on the brane 
hence on the bulk.
Then we can relate it to the 
parameter in RN black hole solution by considering the
full solution to the equation of motion, 
\begin{equation}
R_{mn}-\frac{1}{2}g_{mn}R+g_{mn}\Lambda
=
\kappa^2T_{mn},
\label{eq_motion_bh}
\end{equation}
where energy-momentum tensor $T_{mn}(x)$ is given by
\begin{equation}
T_{mn}
=\frac{1}{e^2}
\bigg(
{\cal F}_{mk}{\cal F}_{nl}g^{kl}
-\frac{1}{4}g_{mn}{\cal F}_{kl}{\cal F}^{kl}
\bigg).
\end{equation}
An equation of motion for the gauge field 
${\cal A}_m(x)$ gives Maxwell equation, 
\begin{equation}
 \nabla_m{\cal F}^{mn}
=\frac{1}{\sqrt{-g}}\del_m
\Big(\sqrt{-g}g^{mk}g^{nl}
\big(\del_k{\cal A}_l-\del_l{\cal A}_k\big)\Big)=0.
\label{maxwell_eq_0}
\end{equation} 
Here we assumed that there is no electromagnetic source outside 
the black hole. 
One can confirm that the following metric and gauge potential
satisfy the equations of motion (\ref{eq_motion_bh}) and 
(\ref{maxwell_eq_0}),
\begin{subequations}
\begin{eqnarray}
\dd s^2
&=&
\frac{r^2}{l^2}
\bigg(
-f(r)(\dd t)^2+\sum_{i=1}^3(\dd x^i)^2
\bigg)
+\frac{l^2}{r^2f(r)}(\dd r)^2,
\label{rnads}
\\
{\cal A}_t
&=&
-\frac{Q}{r^2}+\mu,
\label{rnads_1}
\end{eqnarray}
\end{subequations}

\vspace*{-7mm}
\noindent
with
$$
f(r)
=
1-\frac{ml^2}{r^4}+\frac{q^2l^2}{r^6},
\qquad
\Lambda
=
-\frac{6}{l^2}, \qquad
$$
if and only if $q$ is related to the $Q$ by
\begin{equation}
 e^2=\frac{2Q^2}{3q^2}\kappa^2. 
\end{equation}
It should be noted that 
a ratio of the gauge coupling constant $e^2$ to 
the gravitation constant $\kappa^2$ is 
\begin{equation}
\frac{e^2}{\kappa^2}
=\frac{N_c}{N_f}l^{-2}.
\end{equation}
Since the gauge potential ${\cal A}_t(x)$ must vanish at the horizon, 
the charge $Q$ and the chemical potential $\mu$ are related\footnote{
The chemical potential $\mu$ can be expressed by using gauge invariant 
quantity as 
$$
\mu=\!\int^\infty_{r_+}\!\!\dd r {\cal F}_{rt}={\cal A}_t(\infty), 
$$
where $r_+$ and $\infty$ represent the horizon and the boundary, 
respectively. 
This definition gives thermodynamic relations consistently.  
}.  
The parameters $m$ and $q$ are 
the mass and charge of AdS space, respectively.
This is nothing but Reissner-Nordstr{\"o}m-Anti-deSitter 
(RN-AdS) background
in which we are interested throughout this paper.

The horizons of RN-AdS black hole are located
at the zero for $f(r)$\footnote{
In order to define the horizon, the charge $q$ must satisfy
a relation $q^4\le 4m^3l^2/27$.
},
\begin{equation}
f(r)
=
1-\frac{ml^2}{r^4}+\frac{q^2l^2}{r^6}
=
\frac{1}{r^6}
\Big(r^2-r_+^2\Big)
\Big(r^2-r_-^2\Big)
\Big(r^2-r_0^2\Big), 
\label{metric} 
\end{equation}
where their explicit forms of the horizon radiuses are given by
\begin{subequations}
\begin{eqnarray}
r_+^2
&=&
\left(
\frac{m}{3q^2}
\Bigg(
1+2\cos\bigg(\frac{\theta}{3}+\frac{4}{3}\pi\bigg)
\Bigg)
\right)^{-1},
\label{r+}
\\
r^2_-
&=&
\left(
\frac{m}{3q^2}
\Bigg(
1+2\cos\bigg(\frac{\theta}{3}\bigg)
\Bigg)
\right)^{-1},
\label{r-}
\\
r_0^2
&=&
\left(
\frac{m}{3q^2}
\Bigg(
1+2\cos
\bigg(
\frac{\theta}{3}
+\frac{2}{3}\pi
\bigg)
\Bigg)
\right)^{-1},
\label{r0}
\end{eqnarray}
\end{subequations}

\vspace*{-7mm}
\noindent
with
$$
\theta
=
\arctan
\Bigg(
\frac{3\sqrt{3}q^2\sqrt{\displaystyle 4m^3l^2-27q^4}}{2m^3l^2-27q^4}
\Bigg),
$$
and satisfy a relation $r^2_++r^2_-=-r^2_0$.
The positions expressed by $r_+$ and $r_-$ correspond to the outer
and the inner horizon, respectively. 
It is useful to notice that the charge 
$q$ can be expressed in terms of $\theta$ and $m$ by
$$
q^4=\frac{4m^3l^2}{27}\sin^2\left(\frac{\theta}{2}\right).
$$
The outer horizon takes a value in
$$
\sqrt{\frac{m}{3}}l
\le r_+^2 \le \sqrt{m}l,
$$
where the upper bound and the lower bound correspond to the case for 
$q=0$ and the extremal case, respectively.  

We shall give various thermodynamic quantities of RN-AdS black
hole~\cite{myers, s}.
The temperature is defined from the conical singularity free
condition around the horizon $r_+$,
\begin{equation}
T
=
\frac{r_+^2f'(r_+)}{4\pi l^2}
=
\frac{r_+}{\pi l^2}
\bigg(
1-\frac{1}{2}\frac{q^2l^2}{ r_+^6}
\bigg)
\equiv
\frac{1}{2\pi b}
\Big(
1-\frac{a}{2}  
\Big),
\quad (>0),
\label{temp}
\end{equation}
where we defined the parameters $a$ and $b$ by
\begin{equation}
a\equiv\frac{q^2l^2}{r_+^6}, \qquad
b\equiv\frac{l^2}{2r_+}. 
\end{equation}
In the limit $q\rightarrow 0$,  these parameters go to 
$$
a\rightarrow 0,  \qquad 
b\rightarrow  \frac{l^{3/2}}{2m^{1/4}},
$$
and the temperature becomes 
$$
T\rightarrow T_0
= \frac{m^{1/4}}{\pi l^{3/2}}.
$$
The entropy density $s$, the energy density $\epsilon$, the
pressure $p$, the chemical potential $\mu$ and the density of physical 
charge $\rho$ can be also computed as
\begin{eqnarray}
s
&=&
\frac{2\pi r_+^3}{\kappa^2l^3}
=\frac{\pi l^3}{4b^3\kappa^2},
\label{entropy}
\\
\epsilon
&=&
\frac{3m}{2\kappa^2l^3}
=\frac{3l^3}{32b^4\kappa^2}\Big(1+a\Big), 
\label{energy}
\\
p
&=&
\frac{\epsilon}{3}, 
\label{pressure} 
\\
\mu
&=&
\frac{Q}{r_+^2}=\frac{4b^2Q}{l^4}, 
\label{chemical_potential} 
\\
\rho
&=&
\frac{2Q}{e^2l^3}=\frac{l}{e^2}\frac{\mu}{2b^2}.
\label{charge_density}
\end{eqnarray}

In order to obtain a well-defined boundary term from 
the gravitational part, 
we have to add the Gibbons-Hawking term into the action, 
which is given by
\begin{equation}
 S_{\rm GH} = \frac{1}{\kappa^2}\!\int\! \dd^4x \sqrt{-g^{(4)}} K , 
\label{gh}
\end{equation}
where integration is taken on the boundary of the AdS space. 
The four-dimensional metric $g^{(4)}_{\mu\nu}(x)$ 
is the induced metric on the boundary and $K(x)$ 
is the extrinsic curvature. 
We also need to add counter terms to regularize the action~\cite{bk}, 
\begin{equation}
 S_{\rm ct} = \frac{1}{\kappa^2}\!\int\! \dd^4x \sqrt{-g^{(4)}}
  \left(\frac{3}{l}-\frac{l}{4}R^{(4)}\right) . 
\label{ct}
\end{equation}
%

\section{Perturbations in RN-AdS Background}\label{sec:Perturb}

In RN-AdS background, we study small 
perturbations of the metric $g_{mn}(x)$ and the gauge 
field ${\cal A}_m(x)$,  
\begin{equation}
\begin{array}{rcl}
g_{mn}
&\equiv&
g^{(0)}_{mn}+h_{mn}, 
\\
{\cal A}_m
&\equiv&
A_m^{(0)}+A_m,  
\end{array}
\end{equation}
where the background metric $g^{(0)}_{mn}(x)$ and 
the background gauge field $A^{(0)}_m(x)$ 
are given in (\ref{rnads}) and (\ref{rnads_1}), 
respectively. 
In the metric perturbation, one can define an inverse metric as
$$
g^{mn}=g^{(0)mn}-h^{mn} + h^{ml}h_l{}^n+{\cal O}(h^3),
$$
and raise and lower indices by using the background metric
$g_{mn}^{(0)}(x)$ and $g^{(0)mn}(x)$.

Let us now consider a linearized theory 
of the symmetric tensor field $h_{mn}(x)$ 
and the vector field $A_m(x)$  
propagating in RN-AdS background.
We shall work in the $h_{rm}(x)=0$ and $A_r(x)=0$ gauges and 
use the Fourier
decomposition
\begin{eqnarray*}
h_{\mu\nu}(t, z, r)
&=&
\!\int\!\frac{\dd^4k}{(2\pi)^4}
\ \mbox{e}^{-i\omega t+ikz}h_{\mu\nu}(k, r), 
\\ 
A_\mu(t, z, r)
&=&
\!\int\!\frac{\dd^4k}{(2\pi)^4}
\ \mbox{e}^{-i\omega t+ikz}
A_\mu(k, r), 
\end{eqnarray*}
where we choose the momenta which are along the $z$-direction.  
In this case,
one can categorize the metric perturbations to the following three
types by using the spin under the $SO(2)$ rotation 
in $(x, y)$-plane~\cite{pss}:
\begin{itemize}
\item vector type: \
$h_{tx}\ne0$, \ $h_{zx}\ne0$, \ ${\mbox{(others)}}=0$
\\
\hspace*{23mm}
$\Big($equivalently, $h_{ty}\ne0$, \ $h_{zy}\ne0$, \
${\mbox{(others)}}=0$$\Big)$
\item tensor type: \
$h_{xy}\ne0$, \  ${\mbox{(others)}}=0$
\\
\hspace*{23mm}
$\Big($equivalently, $h_{xx}=-h_{yy}\ne0$, \
${\mbox{(others)}}=0$$\Big)$
\item scalar type: \
$h_{tt}\ne0$, \ $h_{tz}\ne0$, \ $h_{xx}=h_{yy}\ne0$, and $h_{zz}\ne0$,
\ $\mbox{(others)}=0$
\end{itemize}
First two types of the perturbations were studied in~\cite{gmsst}. 
We list the result in Appendix A. 
In this paper we consider the scalar type perturbation. 

\subsection{Linearized equations of motion}

From explicit calculation, one can show that $t$ and $z$-components 
of the gauge field $A_\mu(x)$ could participate in the linearized 
perturbative equations of motion. 
Thus independent variables are 
$$
h_{tt}(x), \quad  
h_{tz}(x), \quad
h_{xx}(x)=h_{yy}(x), \quad
h_{zz}(x), 
$$
$$
A_{t}(x), \quad
A_{z}(x). 
$$

In the hydrodynamic regime, 
it is standard to introduce new dimensionless coordinate 
$u=r^2_+/r^2$ which is normalized by the outer horizon.   
In this coordinate system, the horizon and the boundary are 
located at $u=1$ and $u=0$, respectively. 
We also define new field variables 
\begin{eqnarray*}
h^t_t
&=&
g^{(0)tt}h_{tt}
=-\frac{l^2u}{r_+^2f}h_{tt}, 
\\
h^z_t
&=&
g^{(0)zz}h_{zt}
=\frac{l^2u}{r_+^2}h_{zt}, 
\\
h^x_x
&=&
g^{(0)xx}h_{xx}
=
\frac{l^2u}{r_+^2}h_{xx},
\\
h^z_z
&=&
g^{(0)zz}h_{zz}
=\frac{l^2u}{r_+^2}h_{zz}, 
\\
B_\mu
&\equiv&
\frac{A_\mu}{\mu}=\frac{l^4}{4Qb^2}A_\mu, 
\end{eqnarray*}
where $\mu$ is the chemical potential given by (\ref{chemical_potential}). 
Nontrivial equations in the Einstein equation (\ref{action_bh}) 
appear from $(t, t)$, $(t, u)$, $(t, z)$, 
$(u, u)$, $(u, z)$, $(x, x)$ and $(z, z)$ components, respectively: 
\begin{subequations}
\begin{eqnarray}
0
&=&
{h^t_t}''
+\frac{(u^{-1}f)'}{u^{-1}f}
\Big(\frac{3}{2}{h^t_t}'+{h^x_x}'+\frac{1}{2}{h^z_z}'
\Big)
-\frac{b^2k^2}{uf}h^t_t
\nonumber 
\\
&&
+\frac{2b^2}{uf^2}
\Big(
\omega^2h^x_x
+\frac{1}{2}\omega^2h^z_z
+\omega kh^z_t
\Big)
+2a\frac{u}{f}h^t_t
+4a\frac{u}{f}B_t', 
\label{eq_tt_1}
\\
0
&=&
\omega
\Bigg(
2{h^x_x}'+{h^z_z}'
-\frac{f'}{f}
\Big(h^x_x+\frac{1}{2}h^z_z
\Big)
\Bigg)
+k
\Big(
{h^z_t}'-\frac{f'}{f}h^z_t
\Big), 
\label{eq_tr_1}
\\
0
&=&
{h^z_t}''
-\frac{1}{u}{h^z_t}'
+\frac{2b^2\omega k}{uf}h^x_x
-3auB_z', 
\label{eq_tz_1}
\\
0
&=&
{h^t_t}''
+2{h^x_x}''
+{h^z_z}''
+\frac{f'}{f}
\Big(
\frac{3}{2}{h^t_t}'
+{h^x_x}'
+\frac{1}{2}{h^z_z}'
\Big)
+2a\frac{u}{f}h^t_t
+4a\frac{u}{f}B_t', 
\label{eq_rr_1}
\\
0
&=&
k{h^t_t}'
+2k{h^x_x}'
-\frac{\omega}{f}{h^z_t}'
+\frac{kf'}{2f}h^t_t
+3a\frac{u}{f}
\Big(kB_t+\omega B_z
\Big), 
\label{eq_rz_1}
\\
0
&=&
{h^x_x}''
+\frac{(u^{-2}f)'}{u^{-2}f}{h^x_x}'
-\frac{1}{2u}
\Big({h^t_t}'+{h^z_z}'
\Big)
+\frac{b^2}{uf^2}
\Big(
\omega^2-k^2f
\Big)h^x_x
\nonumber 
\\
&&
-a\frac{u}{f}h^t_t
-2a\frac{u}{f}B_t', 
\label{eq_xx_1}
\\
0
&=&
{h^z_z}''
+\frac{(u^{-\frac{3}{2}}f)'}{u^{-\frac{3}{2}}f}{h^z_z}'
-\frac{1}{u}
\Big(\frac{1}{2}{h^t_t}'+{h^x_x}'
\Big)
+\frac{b^2}{uf^2}
\Big(
\omega^2h^z_z
+2\omega kh^z_t
-k^2fh^t_t
-2k^2fh^x_x
\Big)
\nonumber 
\\
&&
-a\frac{u}{f}h^t_t
-2a\frac{u}{f}B_t', 
\label{eq_zz_1}
\end{eqnarray}
\end{subequations}

\vspace*{-7mm}
\noindent
with 
$$
f(u)=(1-u)(1+u-au^2), 
$$
where the prime implies the derivative with respect to $u$. 
On the other hand, in the Maxwell equation (\ref{maxwell_eq_0}), 
$t$, $u$ and $z$-components give nontrivial contributions 
\begin{subequations}
\begin{eqnarray}
0
&=&
B_t''
-\frac{b^2}{uf}
\Big(
k^2B_t+k\omega B_z
\Big)
+\frac{1}{2}
\Big(
{h^t_t}'-2{h^x_x}'
-{h^z_z}'
\Big), 
\label{eq_at_1}
\\
0
&=&
\omega B_t'
+kf B_z'
+\frac{\omega}{2}
\Big(
h^t_t-2h^x_x-h^z_z
\Big)
-kh^z_t, 
\label{eq_ar_1}
\\
0
&=&
B_z''
+\frac{f'}{f}B_z'
+\frac{b^2}{uf^2}
\Big(
\omega^2B_z+\omega kB_t
\Big)
-\frac{1}{f}
{h^z_t}'.  
\label{eq_az_1}
\end{eqnarray}
\end{subequations}

\vspace*{-7mm}
\noindent
The equations (\ref{eq_at_1}) and (\ref{eq_ar_1}) imply
(\ref{eq_az_1}). 
In the set of equations for the metric 
perturbation (\ref{eq_tt_1})-(\ref{eq_zz_1}), together with 
(\ref{eq_at_1}) and (\ref{eq_ar_1}), 
the following four independent relations are obtained: 
\begin{subequations}
\begin{eqnarray}
{h^x_x}'
&=&
\frac{3(\omega^2-k^2f)+k^2uf'}{2k^2(3f-uf')}{h^t_t}'
+\frac{2b^2\omega^2}{f(3f-uf')}h^x_x
-\frac{f'(3f-uf')-4b^2\omega^2}{4f(3f-uf')}h^t_t
\nonumber 
\\
&&
+\frac{\omega\Big(f'(3f-uf')-4b^2\omega^2\Big)}{2k^2f^2(3f-uf')}
\Big(\omega h^x_x+\frac{\omega}{2}h^z_z+kh^z_t\Big)
\nonumber 
\\
&&
+\frac{3a\omega^2u^2}{2k^2f(3f-uf')}
\Big(h^t_t+2B_t'\Big)
-\frac{3au}{2kf}\Big(kB_t+\omega B_z\Big), 
\label{h_xx_p}
\\
{h^z_z}'
&=&
-\frac{3\omega^2+k^2uf'}{k^2(3f-uf')}{h^t_t}'
-\frac{2b^2k^2}{3f-uf'}\Big(h^t_t+2h^x_x\Big)
+\frac{2b^2}{f(3f-uf')}
\Big(\omega^2(2h^x_x+h^z_z)+2\omega kh^z_t\Big)
\nonumber 
\\
&&
+\frac{1}{2f(3f-uf')}
\Bigg(\Big(f'(3f-uf')-4b^2\omega^2\Big)h^t_t-8b^2\omega^2 h^x_x
\Bigg)
\nonumber 
\\
&&
-\frac{\omega\Big(f'(3f-uf')-4b^2\omega^2\Big)}{k^2f^2(3f-uf')}
\Big(\omega h^x_x+\frac{\omega}{2}h^z_z+kh^z_t\Big)
\nonumber 
\\
&&
-\frac{3au^2(\omega^2+k^2f)}{k^2f(3f-uf')}
\Big(h^t_t+2B_t'\Big)
+\frac{3au}{kf}\Big(kB_t+\omega B_z\Big), 
\label{h_zz_p}
\\
{h^z_t}'
&=&
\frac{3\omega f}{k(3f-uf')}{h^t_t}'
+\frac{2b^2\omega k}{3f-uf'}
\Big(h^t_t+2h^x_x\Big)
+\frac{f'(3f-uf')-4b^2\omega^2}{kf(3f-uf')}
\Big(\omega h^x_x+\frac{\omega}{2}h^z_z+kh^z_t\Big)
\nonumber 
\\
&&
+\frac{3a\omega u^2}{k(3f-uf')}
\Big(h^t_t+2B_t'\Big), 
\label{h_zt_p}
\\
0
&=&
{h^t_t}''
+\frac{1}{2uf(3f-uf')}
\Bigg\{
-3(f-uf')(2f-uf'){h^t_t}'
\nonumber 
\\
&&
\hspace*{41mm}
+4b^2
\Bigg(
-k^2fh^t_t
+\Big(2\omega^2+(f-uf')k^2\Big)h^x_x
+\omega^2h^z_z
+2\omega kh^z_t
\Bigg)
\nonumber 
\\
&&
\hspace*{41mm}
+au^2(15f-7uf')\Big(h^t_t+2B_t'\Big)
\Bigg\}. 
\label{h_tt_pp}
\end{eqnarray}
\end{subequations}

\vspace*{-6mm}
\noindent
The equations of motion (\ref{eq_tt_1})-(\ref{eq_zz_1}) can 
be derived by using the above relations.  
Taking the limit $q\rightarrow 0$, the relations 
(\ref{h_xx_p})-(\ref{h_tt_pp}) coincide with the result in~\cite{ks}. 

\subsection{Surface terms}

Before solving the equations of motion, 
we shall give a surface action in oder to obtain Green functions.  
By using the equations of motion, bilinear parts of 
on-shell action (\ref{action_bh}) reduce to surface terms: 
\begin{eqnarray}
S_0=\frac{l^3}{32\kappa^2b^4}\!\int\!\frac{\dd^4k}{(2\pi)^4}
\Bigg\{
&&
+\frac{f}{u}h^t_t{h^t_t}'
+\frac{f}{u}h^x_x{h^x_x}'
+\frac{f}{u}h^z_z{h^z_z}'
-\frac{3}{u}h^z_t{h^z_t}'
\nonumber 
\\
&&
-\frac{f}{u}h^x_x{h^t_t}' 
-\frac{f}{2u}h^z_z{h^t_t}'
-\frac{f}{u}h^t_t{h^x_x}'
\nonumber
\\
&&
-\frac{f}{u}h^z_z{h^x_x}'
-\frac{f}{2u}h^t_t{h^z_z}'
-\frac{f}{u}h^x_x{h^z_z}'
\nonumber 
\\
&&
+\frac{uf'-f}{4u^2}\left(h^t_t\right)^2
-\frac{f}{4u^2}\left(h_z^z\right)^2 
+\frac{uf'+f}{u^2f}\left(h^z_t\right)^2
\nonumber
\\
&&
-\frac{uf'-2f}{2u^2}h^t_t h_x^x
-\frac{uf'-2f}{4u^2}h^t_t h_z^z
+\frac{f}{u^2} h_x^x h_z^z
\nonumber 
\\
&&
+3a 
\Big(
B_tB_t'-fB_zB_z'
+\frac{1}{2}B_th^t_t
+B_zh^z_t
-B_th^x_x
-\frac{1}{2}B_th^z_z
\Big)
\Bigg\}\Bigg|_{u=0}.  
\nonumber 
\\
\label{surfaceterm}
\end{eqnarray}
Relevant terms of the Gibbons-Hawking term (\ref{gh}) are  
explicitly given by 
\begin{eqnarray}
S_{\rm GH}=\frac{l^3}{32\kappa^2b^4}\!\int\!\frac{\dd^4k}{(2\pi)^4}
\Bigg\{
&&
-\frac{f}{u}h^t_t{h^t_t}'
-\frac{f}{u}h^z_z{h^z_z}'
+\frac{4}{u}h^z_t{h^z_t}'
\nonumber 
\\
&&
+\frac{2f}{u}h^x_x{h^t_t}' 
+\frac{f}{u}h^z_z{h^t_t}'
+\frac{2f}{u}h^t_t{h^x_x}'
\nonumber
\\
&&
+\frac{2f}{u}h^z_z{h^x_x}'
+\frac{f}{u}h^t_t{h^z_z}'
+\frac{2f}{u}h^x_x{h^z_z}'
\nonumber 
\\
&&
-\frac{uf'-4f}{4u^2}\left(h^t_t\right)^2
-\frac{uf'-4f}{4u^2}\left(h_z^z\right)^2 
-\frac{uf'+4f}{u^2f}\left(h^z_t\right)^2
\nonumber
\\
&&
+\frac{uf'-4f}{u^2}h^t_t h_x^x
+\frac{uf'-4f}{2u^2}h^t_t h_z^z
+\frac{uf'-4f}{u^2} h_x^x h_z^z
\Bigg\}. 
\label{gh_01} 
\end{eqnarray}
The counter term (\ref{ct}) also can be evaluated as   
\begin{eqnarray}
S_{\rm ct}
=\frac{3l^3}{32\kappa^2b^4}\!\int\!\frac{\dd^4k}{(2\pi)^4}
\sqrt{f}\Bigg\{
&&
-\frac{1}{4u^2}\left(h^t_t\right)^2
-\frac{1}{4u^2}\left(h_z^z\right)^2 
+\frac{1}{u^2f}\left(h^z_t\right)^2
\nonumber
\\
&&
+\frac{1}{u^2}h^t_t h_x^x
+\frac{1}{2u^2}h^t_t h_z^z
+\frac{1}{u^2} h_x^x h_z^z
\Bigg\}, 
\label{ct_01}
\end{eqnarray}
up to ${\cal O}(\omega^2, k^2, \omega k)$. 

\section{Pole Structures and Transport Coefficients from \\
Hydrodynamics}

We now look for solutions of our set of equations 
(\ref{h_xx_p})-(\ref{h_tt_pp}), and (\ref{eq_at_1}) and (\ref{eq_ar_1}).  
We will consider these equations of motion in
low frequency limit so-called hydrodynamic regime.  
In this regime we could obtain the sound velocity, 
the diffusion constant for $U(1)$ charge and the electrical conductivity 
from retarded Green functions. 

\subsection{Master variables}

By using master variables derived by Kodama and Ishibashi in~\cite{ki}, 
the following field $\Phi(u)$ is introduced:
\begin{equation}
\Phi
\equiv 
\frac{1}{4u^{3/4}(4b^2k^2-3f')}
\Bigg(
\Big(4b^2k^2-3f'\Big)h^x_x
+2f\Big(2{h^x_x}'+{h^z_z}'\Big)
\Bigg).
\end{equation}
For the gauge field, the corresponding variable is given by  
\begin{equation}
{\cal A}
\equiv
2a
\Big(
-h^t_t+3h^x_x-2{B_t}'
\Big). 
\end{equation}
In terms of these new variables $\Phi(u)$ and ${\cal A}(u)$, 
Einstein equations (\ref{h_xx_p})-(\ref{h_tt_pp}) and 
Maxwell equations (\ref{eq_at_1}) and (\ref{eq_ar_1}) can 
be combined as follows: 
\begin{subequations}
\begin{eqnarray}
0
&=&
(u^{1/2}f\Phi')'
\nonumber
\\
&&
-\frac{1}{16u^{3/2}f(4b^2k^2-3f')^2}
\Bigg\{
-16b^2\omega^2u(4b^2k^2-3f')^2
\nonumber 
\\
&&
\hspace*{47mm}
+f^2
\Big(
16(-b^4k^4+108ab^2k^2u^2 +162a^2u^4)
\nonumber 
\\
&&
\hspace*{57mm}
+27f'(8b^2k^2+16au^2+5f')
\Big)
\nonumber 
\\
&&
\hspace*{47mm}
+4uf(4b^2k^2-3f')
\Big(
16b^2k^2(b^2k^2+3au^2)
\nonumber 
\\
&&
\hspace*{83mm}
+f'(8b^2k^2+36au^2+9f')
\Big)
\Bigg\}\Phi
\nonumber 
\\
&&
\hspace*{7mm}
+\frac{1}{8u^{1/4}(4b^2k^2-3f')^2}
\Bigg\{
3f\Big(4b^2k^2+3f'+18au^2\Big)
+u\Big(16b^4k^4-9(f')^2\Big)
\Bigg\}{\cal A}, 
\nonumber 
\\
\label{master_0_phi}
\\
0
&=&
(uf{\cal A}')'
\nonumber
\\
&&
+\frac{1}{f(4b^2k^2-3f')}
\Bigg\{
b^2\Big(4b^2k^2-3f'\Big)\Big(\omega^2-k^2f\Big)
-18auf^2
 \Bigg\}{\cal A}
 \nonumber 
 \\
&&
-48au^{3/4}f\Phi'
\nonumber 
\\
&&
+\frac{4a}{u^{1/4}(4b^2k^2-3f')}
\Bigg\{
32b^4k^4u+12f\Big(b^2k^2+9au^2\Big)
+3f'\Big(-8b^2k^2u+9f\Big)
\Bigg\}\Phi.
\nonumber 
\label{master_0_a}
\\
\end{eqnarray}
\end{subequations}

\vspace*{-6mm}
\noindent
Next we would like to try to obtain decoupled equations from the equations 
(\ref{master_0_phi}) and (\ref{master_0_a}).  
This will be done by introducing the following linear combinations of 
the variables: 
\begin{equation}
\Phi_{\pm}
\equiv
\alpha_{\pm}\Phi+\beta{\cal A}, 
\end{equation}
where the coefficients $\alpha_\pm$ and $\beta$ are  
\begin{eqnarray*}
\alpha_\pm
&=&
C_\pm-3au, 
\\
\beta
&=&
\frac{u^{1/4}}{8}, 
\end{eqnarray*}
with the constants 
$$
C_\pm=(1+a)\pm\sqrt{(1+a)^2+4ab^2k^2}. 
$$
As a result, we can obtain second order ordinary differential 
equations in term of these new variables, 
\begin{eqnarray}
0
&=&
\Phi''_\pm
+\frac{(u^{1/2}f)'}{u^{1/2}f}\Phi'_\pm
+V_\pm\Phi_\pm, 
\label{master_1}
\end{eqnarray}
where potentials $V_\pm(u)$ are given by 
\begin{eqnarray}
V_\pm
&=&
\frac{1}{16u^2f^2(4b^2k^2-3f')^2}
\Bigg\{
16b^2\omega^2u(4b^2k^2-3f')^2
\nonumber 
\\
&&
\hspace*{41mm}
-4uf(4b^2k^2-3f')
\Big(16b^2k^2(b^2k^2-C_\pm u+3au^2)
\nonumber 
\\
&&
\hspace*{78mm}
-4f'(2b^2k^2+3C_\pm u-9au^2)
-3(f')^2
\Big)
\nonumber 
\\
&&
\hspace*{41mm}
+f^2
\Big\{
16
\Big(
b^4k^4+12C_\pm b^2k^2u
\nonumber 
\\
&&
\hspace*{58mm}
-108ab^2k^2u^2
+54C_\pm au^3-162a^2u^4
\Big)
\nonumber 
\\
&&
\hspace*{51mm}
-24f'
\Big(
b^2k^2-6C_\pm u-18au^2
\Big)
+9(f')^2
\Big\}
\Bigg\}. 
\end{eqnarray}

Considering the perturbative expansion with respect to small $\omega$ 
and $k$, 
it might be convenient to introduce new variables 
$\widetilde{\Phi}_\pm(u)$, 
\begin{equation}
\Phi_\pm=H_\pm\widetilde{\Phi}_\pm, 
\end{equation}
where the factors $F_\pm(u)$ are 
$$
H_\pm
=\left\{
\begin{array}{ll}
u^{-3/4} 
&\qquad 
(\mbox{for} \quad \Phi_+) 
\\
\displaystyle
\frac{u^{1/4}}{\displaystyle (1+a)-\frac{3}{2}au}
&\qquad
(\mbox{for} \quad \Phi_-)
\end{array}
\right., 
$$
so that the second order differential equations (\ref{master_1}) are 
reduced to be 
much simpler forms to solve,  
\begin{equation}
0=
\widetilde{\Phi}_\pm''
+\frac{(H_\pm^2u^{1/2}f)'}{H_\pm^2u^{1/2}f}\widetilde{\Phi}'
+\widetilde{V}_\pm\widetilde{\Phi}_\pm, 
\label{master_2}
\end{equation}
where the potential $\widetilde{V}_\pm(u, \omega, k)$ is newly 
defined from the original potential $V_\pm(u, \omega, k)$,  
\begin{equation}
\widetilde{V}_\pm(u, \omega, k)
\equiv
V_\pm(u, \omega, k)-V_\pm(u, 0, 0).
\end{equation}
%

\subsection{Perturbative solutions}

Let us proceed to solve the differential equations. 
First we consider the equation for $\widetilde{\Phi}_+(u)$. 
Following the usual way to solve differential equations, 
we impose a solution as $\widetilde{\Phi}_+(u)=(1-u)^{\nu} F_+(u)$ where 
$F_+(u)$ is a regular function at the horizon $u=1$.   
Plugging this form into the equation of motion, 
one can fix the parameter $\nu$ as $\nu=\pm i\omega/(4\pi T)$ 
where $T$ is the temperature defined by eq. (\ref{temp}). 
We here choose 
\begin{equation}
\nu=-i\frac{\omega}{4\pi T}
\end{equation}
as the incoming wave condition.
We are now in the position to solve the equation of motion 
in the hydrodynamic regime. 
We start by introducing the following series expansion with respect 
to small $\omega$ and $k$: 
\begin{equation}
F_+(u)
=
F_{+0}(u)+\omega F_{+1}(u)+k^2G_{+1}(u)+\omega^2 F_{+2}(u)
+{\cal O}(\omega^3, \ \omega k^2), 
\label{expansion_1} 
\end{equation}
where $F_{+0}(u)$, $F_{+1}(u)$ and $G_{+1}(u)$ are determined by imposing 
suitable boundary conditions. 
The solution can be obtained recursively\footnote{
The derivation of the solution is given in Appendix B.
}.
The result is as follows: 
\begin{subequations}
\begin{eqnarray}
F_{+0}(u)
&=&
C, \quad (\mbox{const.}) 
\\
F_{+1}(u)
&=&
\frac{iCb}{2 (2-a)}
\left\{\log \left(1+u-au^2\right)
-\frac{6 K_1(u)}{\sqrt{1+4a}}
\right\},
\\
G_{+1}(u)
&=&
\frac{2}{3}Cb^2 
\left\{\frac{K_1(u)}{\sqrt{1+4a}}-\frac{1}{(1+a)u}
\right\}, 
\\
F_{+2}(u)
&=&\!\int^u\!\dd u\frac{Cb^2}{(1-u)(1+u-au^2)}
\nonumber 
\\
&&
\hspace*{5mm}
\times
\Biggl\{1-u+\frac{(1-u) (1+au) \log \left(1+u-a u^2\right)}{2 (2-a)^2}
-\frac{3 (1-u) (1+au) K_2(0)}{2 (2-a)^2 \sqrt{1+4a}}
\nonumber 
\\
&&
\hspace*{12mm}
-\frac{(1+a) K_2(1) u}{\sqrt{1+4a}}
+\frac{\Big(3+(5+3a-6a^2+2a^3)u-3au^2\Big)
   K_2(u)}{2 (2-a)^2 \sqrt{1+4a}}\Biggr\}, 
\nonumber
\\ 
\end{eqnarray}
\end{subequations}
where
\begin{eqnarray*}
K_1(u)
&=&\frac{1}{2}\log(1+u-au^2)-\log\left(1-\frac{2au}{1+\sqrt{1+4a}}\right),
\\
K_2(u)
&=&\log\left(\frac{1+\sqrt{1+4a}-2au}{-1+\sqrt{1+4a}+2au}\right). 
\end{eqnarray*}

Next, we shall study the equation of motion for
$\widetilde{\Phi}_-(u)$. 
Assuming again $\widetilde{\Phi}_-(u)=(1-u)^{\nu} F_-(u)$ 
where $F_-(u)$ is a regular function at $u=1$, 
the singularity might be extracted. 
We fix the constant as $\nu=-i\omega/(4\pi T)$ to use  
the incoming wave condition.  
We now impose a perturbative solution as 
\begin{equation}
F_-(u)
=
F_{-0}(u)+\omega F_{-1}(u)+k^2 G_{-1}(u)+\omega^2 
F_{-2}(u)+{\cal O}(\omega^3, \ \omega k^2), 
\label{expansion_2}
\end{equation}
and then we obtain the following result\footnote{
The detail is given in Appendix C.
}:
\begin{subequations}
\begin{eqnarray}
F_{-0}(u)
&=&
\widetilde{C}, \quad (\mbox{const.})
\\
F_{-1}(u)
&=&
\frac{i\widetilde{C}b}{2 (2-a)^2}
\Bigl\{
8(1+a)^2 \log (u)-(2+a) (1+4a) 
\log\Big(1+u-au^2\Big)
\nonumber
\\
&&
\hspace{20mm}
-2 \sqrt{1+4a} (2+5a) K_1(u)
\Bigr\}, 
\\ 
G_{-1}(u)
&=&
\tC b^2 
\Biggl\{-\frac{3a^2u}{2(1+a)^2\Big(\displaystyle 1+a-\frac{3}{2}au\Big)}
-\frac{2 (1+a) (2+a) \log (u)}{(2-a)^2} 
\nonumber\\
&&
\hspace{10mm}
+\frac{(1+a)(2+a) 
\log\Big(1+u-au^2\Big)}{(2-a)^2}+\frac{2 (2+5a+6a^2) K_1(u)}{(2-a)^2 
\sqrt{1+4a}}
\Biggr\}, 
\\
F_{-2}(u)
&=&
\!\int^u\!\dd u
\frac{\widetilde{C}b^2 }{2 (2-a)^4 (1+4a)^{3/2} (1-u) u 
\left(1+u-a u^2\right)}
\nonumber\\
&&
\hspace*{5mm}\times
\Biggl\{8 (2-a)(1+a)^2(1+4a)^{3/2} u (1+u-au^2)\log (u)
\nonumber\\
&&
\hspace*{11mm}
-(2-a) (1+4a)^{3/2} 
\Bigl(
4(1+a)^2
+(2-3a-8a^2)u
+(2+9a+13a^2)u^2
\nonumber\\
&&
\hspace{48mm}
-a (2+a) (1+4a)u^3
\Bigr) 
\log \Big(1+u-a u^2\Big)
\nonumber\\
&&
\hspace*{11mm}
+(1+4a)^2 (2+5a)K_2(0) 
(1-u) 
\Bigl(4(1+a)^2+(2-3a-8a^2)u
\nonumber 
\\
&&
\hspace*{71mm}
-(2-a)au^2
\Bigr)
\nonumber\\
&&
\hspace*{11mm}
-2(1+a)(2-2a+41a^2)K_2(1)\Big(2(1+a)-3au\Big)^2
\nonumber\\
&&
\hspace*{11mm}
-(2-a)
\Bigl(a(1+4a)^2(2+5a)u^3
-(2-a)(1+11a+46a^2+18a^3)u^2
\nonumber\\
&&
\hspace*{29mm}
-\left(2+9a+180a^2+224a^3+24a^4\right)u
\nonumber 
\\
&&
\hspace*{29mm}
-4(1+a)^2 (1-10a-2a^2)
\Bigr)K_2(u)
\Biggr\}.
\end{eqnarray}
\end{subequations}

\vspace*{-6mm}

Using these, we can get  
behaviors for the solutions of $\Phi_+(u)$ and $\Phi_-(u)$ 
around the boundary $u=0$, 
\begin{subequations}
\begin{eqnarray}
 \Phi_+ 
&=& 
\frac{C}{u^{3/4}} \left\{1-\frac{2 k^2 b^2}{3 (1+a) u}
 +\frac{1}{3} \left(k^2+3 \omega^2\right)b^2 u +\cdots\right\}, 
   \\
\Phi_- 
&=& 
\tC u^{1/4} 
\left\{\frac{1}{1+a}+\left(\frac{4 i (1+a) b \omega}{(2-a)^2}
-\frac{2 (2+a) b^2 k^2}{(2-a)^2}+b^2 
 D_- \omega^2\right) \log
   (u)+\cdots\right\}, 
\nonumber 
\\
\end{eqnarray}
\end{subequations}

\vspace*{-6mm}
\noindent
where the constant $D_-$ is 
\begin{eqnarray*}
D_-
&=&
\frac{2}{(2-a)^4 (1+4a)^{3/2}}
\Bigg\{
-27 (2-a) a^2 \sqrt{1+4a}
\nonumber\\
&&
\hspace*{37mm}
+4 (1+4a)^{3/2} (1+a)^3 \log (2-a)
\nonumber 
\\
&&
\hspace*{37mm}
+4 (2-2a+41a^2)(1+a)^2 K_1(1)\Bigg\}. 
\end{eqnarray*}

Let us now consider the integration constants $C$ and $\tilde{C}$. 
These could be estimated in terms of boundary values of the fields
$$
\begin{array}{rclcrclcrclcrclcrcl}
\displaystyle
\lim_{u\rightarrow 0}h^t_t(u)
&=&(h^t_t)^0, 
\displaystyle
\qquad \lim_{u\rightarrow 0}B_t(u)
=(B_t)^0,
\qquad {\mbox{etc}}.   
\end{array}
$$
Using equations of motion, the integration constants 
$C$ and $\tC$ are determined as 
\begin{subequations}
\begin{eqnarray}
 C 
&=& 
 \frac{1}{2 \left(k^2-3 \omega^2\right)}
 \Biggl\{3 a k^2(B_t)^0 +3 a k\omega (B_z)^0 
 \nonumber\\
 &&
 \hspace{25mm}
+(1+a)\Big(-k^2 (h^t_t)^0 +2 k\omega (h^z_t)^0 +\left(k^2-\omega^2\right)
   (h^x_x)^0+\omega^2 (h^z_z)^0\Big)\Biggr\}, 
\nonumber 
\\
\label{cc}
&&
\\
\tC 
&=& 
\frac{(2-a)^2 a b k \Big(k (B_t)^0+\omega (B_z)^0\Big)}
{2D_{\rm p}(\omega, k)}, 
\label{cct}
\end{eqnarray}
\end{subequations}
where
$$
D_{\rm p}(\omega, k)=2 (2+a) b k^2-4 i (1+a) \omega- (2-a)^2 b D_- \omega^2. 
$$
In the equations (\ref{cc}) and (\ref{cct}), 
one can observe the existence of the sound and diffusion poles in the 
complex $\omega$-plane. 

\subsection{Retarded Green functions} 

Let us evaluate the Minkowskian correlators. 
The relevant action is given by the sum of 
three parts (\ref{surfaceterm}), (\ref{gh_01}) and (\ref{ct_01}), 
\begin{eqnarray}
S
&=&
S_0+S_{\rm GH}+S_{\rm ct}
\nonumber 
\\
&=&
\frac{l^3}{32\kappa^2b^4}\!\int\!\frac{\dd^4k}{(2\pi)^4}
\Bigg\{
\frac{1}{u}h^z_t{h^z_t}'
+\frac{f}{u}h^x_x{h^x_x}'
+\frac{f}{u}h^t_t{h^x_x}'
+\frac{f}{2u}h^t_t{h^z_z}'
\nonumber 
\\
&&
\hspace*{30mm}
+\frac{f}{u}h^x_x{h^t_t}' 
+\frac{f}{u}h^x_x{h^z_z}'
+\frac{f}{2u}h^z_z{h^t_t}'
+\frac{f}{u}h^z_z{h^x_x}'
\nonumber 
\\
&&
\hspace*{30mm}
+\frac{3}{4u^2}\Big(f-\sqrt{f}\Big)(h^t_t)^2
+\frac{1}{4u^2}\Big(3f-uf'-3\sqrt{f}\Big)(h^z_z)^2
\nonumber 
\\
&&
\hspace*{30mm}
-\frac{3}{u^2f}\Big(f-\sqrt{f}\Big)(h^z_t)^2
-\frac{1}{2u^2}\Big(6f-uf'-6\sqrt{f}\Big)h^t_th^x_x
\nonumber 
\\
&&
\hspace*{30mm}
-\frac{1}{4u^2}\Big(6f-uf'-6\sqrt{f}\Big)h^t_th^z_z
-\frac{1}{u^2}\Big(3f-uf'-3\sqrt{f}\Big)h^x_xh^z_z
\nonumber 
\\
&&
\hspace*{30mm}
+3a 
\Big(
B_tB_t'-fB_zB_z'
+\frac{1}{2}B_th^t_t
+B_zh^z_t
-B_th^x_x
-\frac{1}{2}B_th^z_z
\Big)
\Bigg\}\Bigg|_{u=0}.  
\nonumber
\\ 
\label{surfaceterm+}
\end{eqnarray}
Using equations of motion and solutions of $\Phi_+(u)$ and $\Phi_-(u)$,  
derivatives of $h$'s and $B$'s can be expressed 
in terms of their boundary values: 
\begin{subequations}
\begin{eqnarray}
\frac{1}{u}{h^t_t}' 
&\to& 
\frac{3}{k^2-3\omega^2}
\Bigg\{
-3 a k 
\Big(k (B_t)^0+\omega(B_z)^0\Big)
\nonumber 
\\
&&
\hspace*{20.5mm}
+(1+a) 
\Big(k^2(h^t_t)^0 -2\omega k(h^z_t)^0 -\omega^2 (2
   (h^x_x)^0+(h^z_z)^0)
\Big)
\Bigg\},
\\
\frac{1}{u}{h^x_x}' 
&\to&
\frac{1}{k^2-3\omega^2} 
\Bigg\{
3 a k 
\Big(k (B_t)^0+\omega (B_z)^0
\Big)
\nonumber 
\\
&&
\hspace*{20mm}
+(1+a) 
\Big(
-k^2(h^t_t)^0 +2 k\omega
(h^z_t)^0 +\omega^2 (2 (h^x_x)^0+(h^z_z)^0)
\Big)\Bigg\},
\\
\frac{1}{u}{h^z_z}' 
&\to& 
\frac{1}{k^2-3\omega^2}
\Bigg\{
3 a k 
\Big(
k (B_t)^0+\omega (B_z)^0
\Big)
\nonumber 
\\
&&
\hspace*{20mm}
+(1+a)
\Big(
-k^2(h^t_t)^0 +2 k\omega
(h^z_t)^0 +\omega^2 (2 (h^x_x)^0+(h^z_z)^0)
\Big)\Bigg\}, 
\\
\frac{1}{u}{h^z_t}' 
&\to&
\frac{1}{k^2-3\omega^2} 
\Bigg\{-9 a \omega 
\Big(k (B_t)^0+\omega (B_z)^0\Big)
\nonumber 
\\
&&
\hspace*{20.5mm}
-(1+a)k 
\Big(
2 k(h^z_t)^0+\omega (-3 (h^t_t)^0+2 (h^x_x)^0+(h^z_z)^0)
\Big)\Bigg\}, 
\\
B'_t 
&\to&
\frac{1}{2(1+a)(k^2-3\omega^2)}
\Bigg\{ 
-9 a k 
\Big(
k (B_t)^0+\omega (B_z)^0
\Big)
\nonumber 
\\
&&
\hspace*{38mm}
+(1+a) 
\Big(
(2 k^2+3\omega^2) (h^t_t)^0
-6k\omega(h^z_t)^0
\nonumber 
\\
&&
\hspace*{56mm}
-3 \omega^2(2(h^x_x)^0
   + (h^z_z)^0)\Big)
\Bigg\}
\nonumber
\\
&&
-\frac{(2-a)^2bk\Big(k (B_t)^0+\omega (B_z)^0\Big)}
{(1+a)D_{\rm p}(\omega, k)}, 
\\
B'_z 
&\to& 
\frac{1}{2(1+a)(k^2-3\omega^2)}
\Bigg\{
9a\omega 
\Big(
k(B_t)^0+\omega (B_z)^0
\Big)
\nonumber 
\\
&&
\hspace*{38mm}
+(1+a) 
\Big(-3
   k\omega (h^t_t)^0 +2  k^2(h^z_t)^0
+k\omega(2(h^x_x)^0+(h^z_z)^0) \Big)
\Bigg\}
\nonumber\\
&&
+\frac{(2-a)^2 b \omega \Big(k (B_t)^0+\omega
   (B_z)^0\Big)}
{(1+a) D_{\rm p}(\omega, k)}. 
\end{eqnarray}
\end{subequations}

\vspace*{-6mm}
\noindent
Substituting these expressions to the surface term (\ref{surfaceterm+}), 
we can read off the Green functions defined by 
(\ref{diff_green_function}).  
Through the counter terms, the singularities around the boundary 
vanish completely. 
The results are listed below in 
Table~\ref{tab:Gh-h}, \ref{tab:Gh-a} and \ref{tab:Ga-a}.
 
%
%
\begin{table}[htbp]
\caption{$\displaystyle G_{**~ **}~/~\left
(\frac{-(1+a)l^3}{128\kappa^2b^4(k^2-3\omega^2)}\right)$. 
}
\label{tab:Gh-h}
\begin{center}
\begin{tabular}{|c||c|c|c|c|}
\hline
&$tt$&$xx$&$zz$&$tz$\\
\hline
\hline
$tt$&\makebox[3cm][c]{$3
\left(5k^2-3\omega^2\right)$}
&\makebox[3cm][c]{$6
\left(k^2+\omega^2\right)$}
&\makebox[3cm][c]{$3
\left(k^2+\omega^2\right)$}
&\makebox[3cm][c]{24$
\left(k^2+\omega^2\right)$}\\
\hline
$xx$&---&\makebox[3cm][c]{$16\omega^2$}
&\makebox[3cm][c]{$2
\left(k^2+\omega^2\right)$}
&\makebox[3cm][c]{$16k\omega$}\\
\hline
$zz$&---&---&\makebox[3cm][c]{$-k^2+7\omega^2$}
&\makebox[3cm][c]{$8k\omega$}\\
\hline
$tz$&---&---&---&\makebox[3cm][c]{$4(k^2+9\omega^2)$}\\
\hline
\end{tabular}
\end{center}
\end{table}
%
%
%
\begin{table}[htbp]
\caption{$\displaystyle G_{**~ *}
~/~\left(\frac{-l\mu}{4e^2b^2(k^2-3\omega^2)}\right)$. 
}
\label{tab:Gh-a}
\begin{center}
\begin{tabular}{|c||c|c|c|c|}
\hline
&$tt$&$xx$&$zz$&$tz$\\
\hline
\hline
$t$&\makebox[3cm][c]{$3k^2$}
&\makebox[3cm][c]{$2k^2$}
&\makebox[3cm][c]{$k^2$}
&\makebox[3cm][c]{$6k\omega$}\\
\hline
$z$&\makebox[3cm][c]{$3k\omega$}
&\makebox[3cm][c]{$2k\omega$}
&\makebox[3cm][c]{$k\omega$}
&\makebox[3cm][c]{$6\omega^2$}\\
\hline
\end{tabular}
\end{center}
\end{table}
%
%
%
\begin{table}[htbp]
\caption{$\displaystyle 
G_{*~ *}~/~\left(
\frac{-l}{4e^2(1+a)b^2}\left(\frac{9a}{k^2-3\omega^2}
+\frac{2(2-a)^2b}{D_{\rm p}(\omega, k)}\right)\right)$. 
}
\label{tab:Ga-a}
\begin{center}
\begin{tabular}{|c||c|c|}
\hline
&$t$&$z$\\
\hline
\hline
$t$&\makebox[3cm][c]{$k^2$}
&\makebox[3cm][c]{$k\omega$}\\
\hline
$z$&\makebox[3cm][c]{---}
&\makebox[3cm][c]{$\omega^2$}\\
\hline
\end{tabular}
\end{center}
\end{table}
%
%

In the final expression we rescaled the gauge 
field $(B_\mu)^0$ to the original 
one $(A_\mu)^0=\displaystyle\frac{4Qb^2}{l^4}(B_\mu)^0$ 
and raised and lowered the indices by 
using the flat Minkowski metric $\eta_{\mu\nu}={\rm diag}(-,+,+,+)$ in 
four-dimensional boundary theory. 
Taking the limit which the charge goes to zero, 
the correlators for energy-momentum tensors coincide with 
the known ones in \cite{pss}. 
In this limit, the correlators for the energy-momentum tensor and the $U(1)$ 
current vanish, while ones for the $U(1)$ currents have no 
sound poles, as we could see in the case of vector type perturbation 
for $(1,0,0)$ R-charged~\cite{ss2} and RN-AdS black holes~\cite{gmsst}.  

Remember  
$l^3/\kappa^2=N_c^2/(4\pi^2)$ and 
it should be noticed that the factor $l/e^2=N_c^2/(16\pi^2)$ 
for the R-charge since $e=2\kappa/l$ in that case, 
while  $l/e^2=N_cN_f/(4\pi^2)$ 
for the brane charge  \cite{s}.

\subsection{Transport coefficients}  

From the obtained Green functions, 
we can observe the value of the speed of sound 
without the medium effect, 
\begin{equation}
v_{\rm s} 
=\frac{1}{\sqrt{3}}. 
\end{equation}
One should notice that there is no effect of the charge on 
the sound velocity. 

We can also find the diffusion pole in 
the current-current correlators. 
The diffusion constant can be read off 
\begin{equation}
D_A
=\frac{(2+a)b}{2(1+a)}. 
\end{equation}
It  should be compared with  
the diffusion constant for gravitation fields $D_H$ 
obtained in  the previous work~\cite{gmsst}, 
$$
D_H=\frac{b}{2(1+a)}, 
$$
so that the relation between them is 
\begin{equation}
D_A=(2+a)D_H.
\end{equation}
In the chargeless case, we can reproduce the result in~\cite{pss}. 

The electrical conductivity $\sigma$ of the medium 
could be also determined by the current-current correlators 
via Kubo formula, 
$$
\sigma 
\equiv
-\lim_{\omega\rightarrow 0}\frac{e_{\rm E}^2}{3\omega}
{\mbox{Im}}\Big(\delta^{ij}G_{ij}(\omega, k=0)\Big),  
$$
where $e_{\rm E}$ is a four-dimensional gauge coupling. 
Together with the result for vector type perturbation~\cite{gmsst}, 
we can obtain 
\begin{eqnarray}
\sigma
&=&
\frac{le_{\rm E}^2(2-a)^2}{24e^2(1+a)^2b}\times 3 
\nonumber 
\\
&=&
\left(\frac{l}{e^2}\right)
\frac{\pi {e_{\rm E}}^2 (2-a)}{2(1+a)^2} T. 
\end{eqnarray}

We can also access to the charge susceptibility $\Xi$ defined by 
\begin{equation}
\Xi\equiv\frac{1}{T}\frac{\langle Q^2\rangle}{\mbox{(volume)}}. 
\label{cs}
\end{equation}
Using Green function $G_{t \ t}(k)$ which might give  
an expectation value of $Q^2$, one can obtain the following 
relation in thermal equilibrium,   
\begin{equation}
\frac{\langle Q^2\rangle}{\mbox{(volume)}}
=
\int
\frac{\dd\omega}{2\pi}
\Big(-\mbox{Im}\Big(G_{t \ t}(\omega, k\rightarrow 0)\Big)\Big)
n_{\rm b}(\omega), 
\end{equation}
where $n_{\rm b}(\omega)=\frac{1}{\mbox{e}^{\omega/T}-1}$ is 
Bose-Einstein distribution function~\cite{hkmsy}.  
From Table 3, we can see 
$$
-\mbox{Im}\Big(G_{t \ t}(\omega, k)\Big)
=
\frac{4\pi^2lT^2}{e^2(1+a)(2+a)}
\Bigg(
\frac{\omega D_Ak^2}{\omega^2+(D_Ak^2)^2}
\Bigg). 
$$ 
It should be noted that 
a quantity $D_Ak^2/\Big(\omega^2+(D_Ak^2)^2\Big)$ approaches to  
$2\pi\delta(\omega)$ for $k\rightarrow 0$ limit.  
Taking the relation (\ref{cs}) into account, we can read off the 
charge susceptibility $\Xi$ as  
\begin{equation}
\Xi
=
\left(\frac{l}{e^2}\right)
\frac{4\pi^2 }{ (1+a)(2+a)} T^2. 
\label{xi}
\end{equation}
We can then observe that Einstein relation 
\begin{equation}
\sigma/({e_{\rm E}}^2\Xi)=D_A
\end{equation}
holds exactly. (See also \cite{mst}.)\footnote{
It is interesting to notice that 
$(\del\rho/\del\mu)_T$ gives a different value for $\Xi$ 
given in eq.(\ref{xi}).
The authors thank J. Mas and J. Shock for pointing this out.
}

In R-charge case, taking the charge-free limit, 
the electric conductivity and the charge susceptibility 
coincide with the results in~\cite{hkmsy}. 

\begin{figure} [htbp]
\begin{minipage}{0.43\hsize}
\begin{center}
\includegraphics*[scale=0.5]{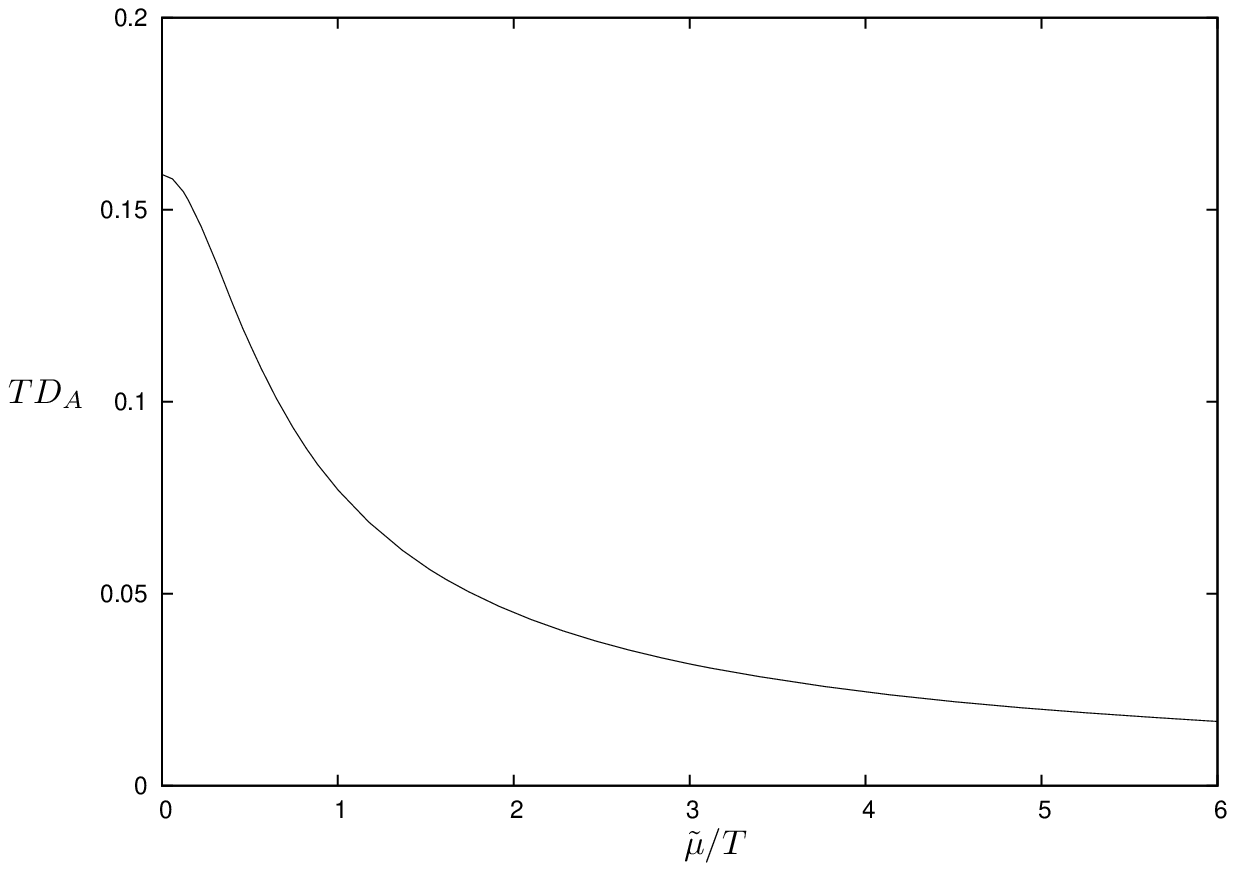}
\end{center}
\caption{$T D_A$ vs. $\tilde\mu/T$}
\label{fig_da}
\end{minipage}
\qquad\quad
\begin{minipage}{0.43\hsize} 
\begin{center}
\includegraphics*[scale=0.5]{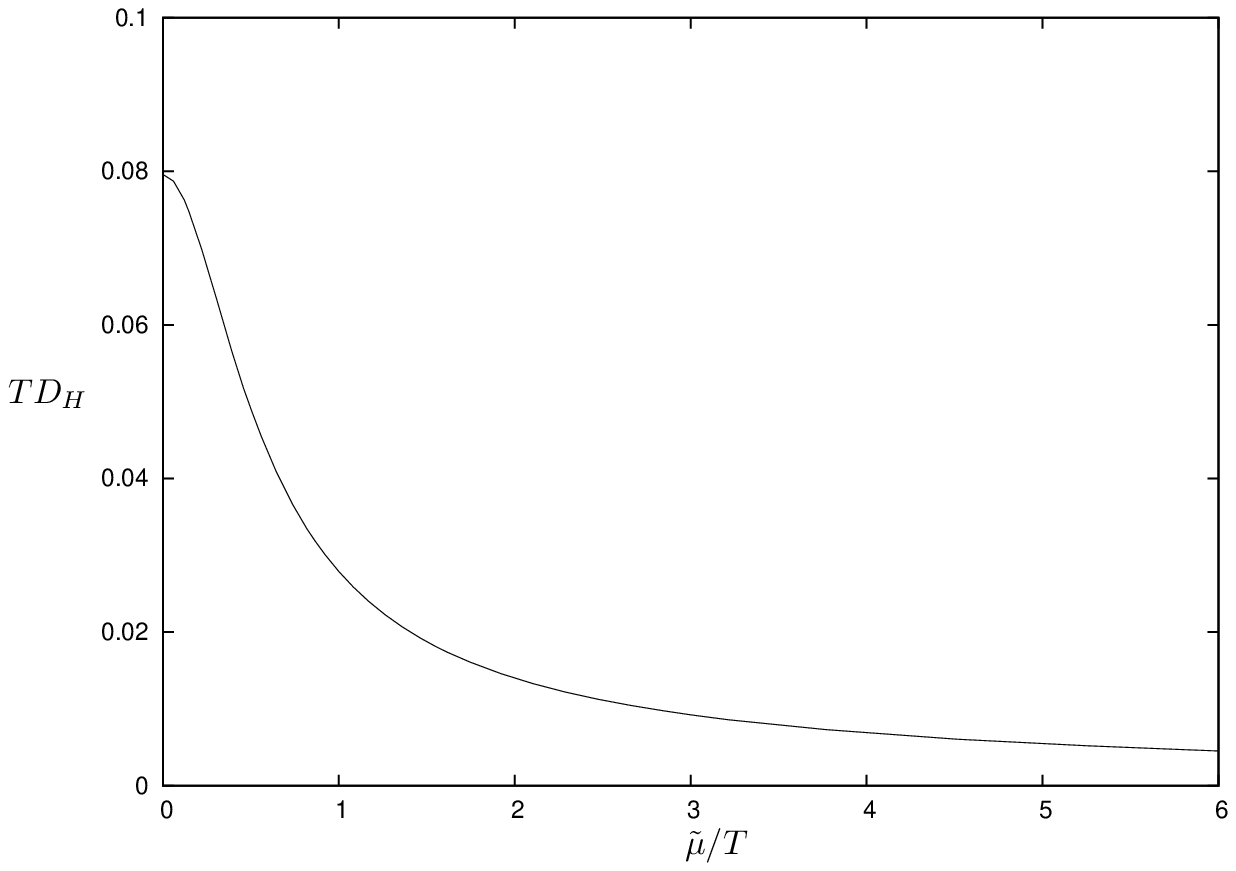}
\end{center}
\caption{$TD_H$ vs. $\tilde\mu/T$}
\label{fig_DH}
\end{minipage}
\end{figure}

\begin{figure} [htbp]
\begin{minipage}{0.43\hsize}
\begin{center}
\includegraphics*[scale=0.5]{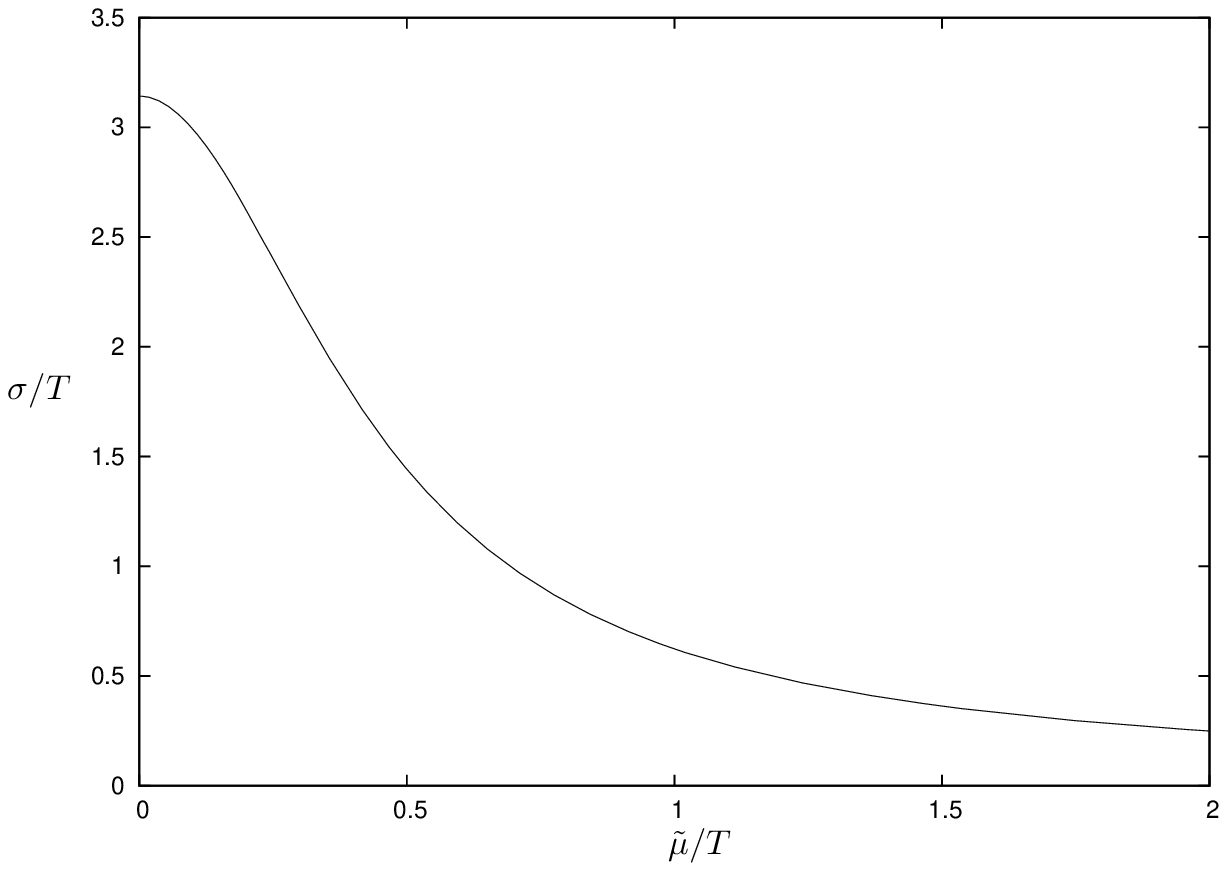}
\end{center}
\caption{$\sigma/(Tle_{\rm E}^2/e^2)$ vs. $\tilde\mu/T$.}
\label{fig_cond}
\end{minipage}
\qquad\quad
\begin{minipage}{0.43\hsize} 
\vspace*{6mm}
\begin{center}
\includegraphics*[scale=0.5]{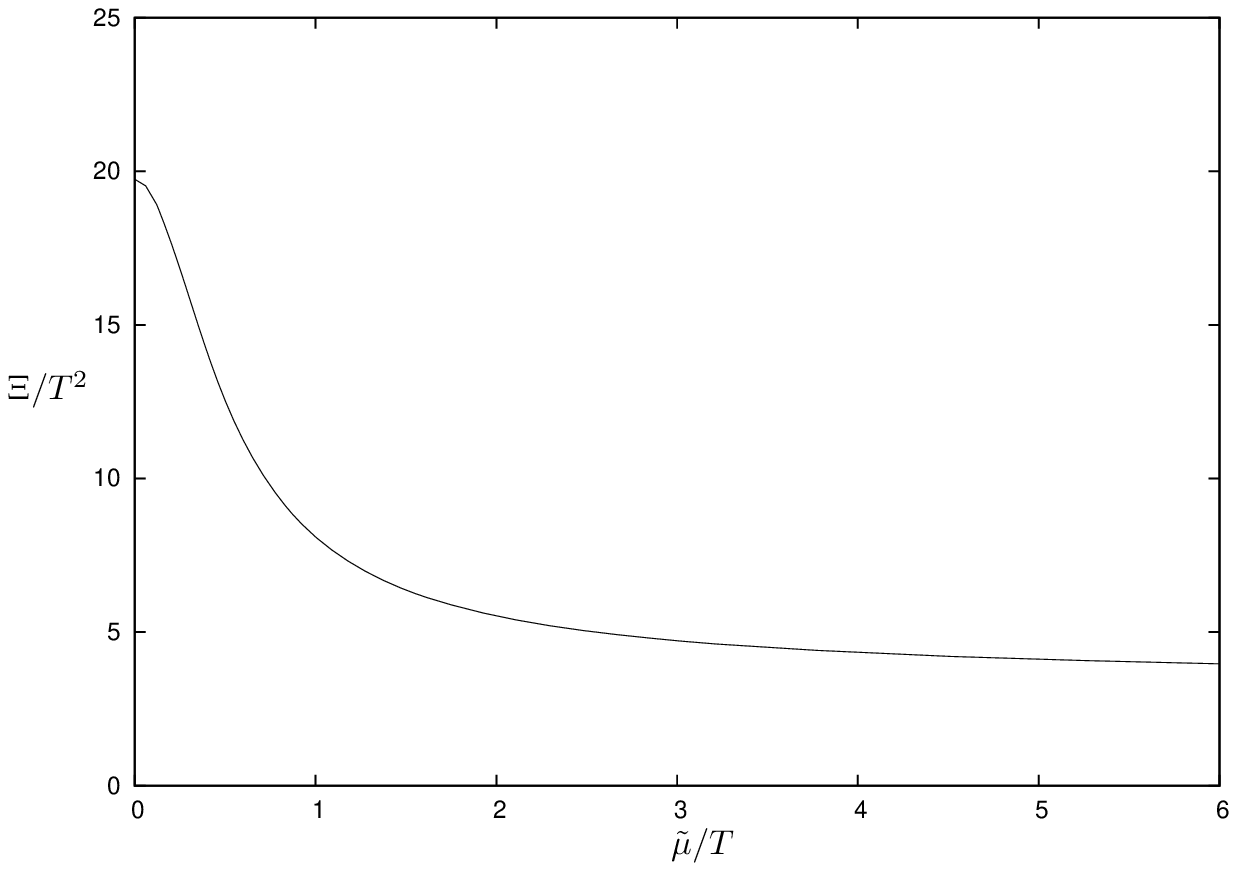}
\end{center}
\caption{$\Xi/(T^2l/e^2)$ vs. $\tilde\mu/T$
: Notice the rapid change between two finite values as $T$ 
runs from 0 to $\infty$}
\label{fig_sus}
\end{minipage}
\end{figure}

It is interesting to express physical constants in terms of the 
boundary variables: the temperature  and the chemical potential. 
In fact,  it is easy to verify that 
\begin{equation}
a=2-\frac{4}{1+\sqrt{1+4(\tilde{\mu}/T)^2}},\qquad 
b=\left(\frac{1}{\pi T}\right)
\frac{1}{1+\sqrt{1+4(\tilde{\mu}/T)^2}}, 
\end{equation}
where 
$\displaystyle{\tilde \mu}\equiv\frac{\kappa}{\sqrt{3}\pi el}\mu$. 
Notice that for the R-charge, 
$\displaystyle\tilde{\mu}=\frac{1}{2\sqrt{3}\pi}\mu$,  
while for the brane charge, 
$\displaystyle\tilde{\mu}=\sqrt{\frac{N_f}{3N_c\pi^2}}\mu$.
The behaviors of the diffusion constants $D_A$ and $D_H$, the electrical
conductivity $\sigma$ and the charge susceptibility $\Xi$ are 
drawn as functions of 
the  $\tilde\mu/T$ in Figure~\ref{fig_da}, \ref{fig_DH},
\ref{fig_cond} and \ref{fig_sus} respectively.  
Notice that for the fixed temperature, 
all of them are decreasing functions of the chemical potential. 
One should notice that there is no upper bound of $\mu/T$ 
for any of these quantities unlike  (1,0,0) charged black hole 
studied in~\cite{ss2}. 
 
It is particularly interesting to observe 
that the charge susceptibility modulo  $T^2$ factor, 
which is an indicator of the degree of freedom, 
shows rapid change between low and high temperature (density)
indicating a mild phase transition.  
Such behavior does not exist for chargeless case. 
See also \cite{jkls}.
 
\section{Conclusions and Discussions}\label{sec:Conclusion}

In this paper, we worked out the decoupling of scalar modes of the 
charged AdS black hole background in $SO(2)$ basis for the mode 
classifications. 
We also perform the hydrodynamic analysis for the holographic 
Quark-Gluon Plasma system. 
Master equations for the decoupled modes are worked out explicitly. 
The sound velocity is not modified by the presence of the charge. 
We calculated the diffusion constants, the charge susceptibility and 
the conductivity as a consequence and observed that Einstein relation holds 
between them. 
These transport coefficients are modified due to the charge effect. 
Interestingly,  the susceptibility modulo  $T^2$ factor, 
which is an indicator of the degree of freedom, shows rapid change 
between low and high temperature (density)
indicating a mild phase transition. 
Such behavior does not exist for chargeless case. 

One can give an explanation of hydrodynamic mode in meson physics. 
In our interpretation, the Maxwell fields are the fluctuations of 
bulk-filling branes, therefore they should be interpreted as
master fields of the mesons. 
Then hydrodynamic modes are lowest lying massless meson spectrum. 
In terms of brane embedding picture, 
this massless-ness is due to the touching of the brane 
on the black hole horizon. 
Near the horizon, the tension of the brane is zero due to 
the metric factor and it can lead to the massless fluctuation. 
Then the massless spectrum can not go far from the horizon 
in radial direction.  
In this picture, hydrodynamic nature is closely related 
to the near horizon behavior of the branes. 
We will discuss the spectrum of meson mode by considering 
the quasinormal mode~\cite{kz} of the vector modes in elsewhere. 

\vspace*{5mm}

\noindent
{\large{\bf Acknowledgments}}

\vspace*{2mm}

We would like to thank X.-H. Ge and F.-W. Shu for useful discussion 
at the early stage of this work and S. Nakamura for stimulating
discussions.  
We especially want to thank J. Mas and J. Shock for pointing out 
interesting points after first version of the paper was uploaded. 
The work of SJS was supported by KOSEF Grant R01-2007-000-10214-0. 
This work is also supported by Korea Research Foundation Grant 
KRF-2007-314-C00052  and  SRC Program of the KOSEF through the 
CQUeST with grant number R11-2005-021.

\vspace*{5mm}

\appendix

\section{Results for the vector and the tensor type perturbations} 

\subsection{Vector type perturbation}

In the vector type perturbation, independent variables are  
$$
h_{tx}(x), \quad
h_{zx}(x), \quad
A_x(x).
$$
One can observe the diffusion constant for the metric perturbation 
$D_H$ as 
\begin{equation}
D_H=\frac{b}{2(1+a)}. 
\end{equation}
We list the retarded Green functions in 
Table~\ref{tab:Gh-h_vector}, \ref{tab:Gh-a_vector} and 
\ref{tab:Ga-a_vector}. 
%
%
\begin{table}[bhtp]
\caption{
$\displaystyle G_{**~ **}~/~\left
(\frac{l^3}{16\kappa^2b^3(i\omega-D_Hk^2)}\right)$. 
}
\label{tab:Gh-h_vector}
\begin{center}
\begin{tabular}{|c||c|c|}
\hline
&$tx$&$zx$
\\
\hline
\hline
$tx$&\makebox[3cm][c]
{$k^2$}
&\makebox[3cm][c]
{$-\omega k$}
\\
\hline
$zx$
&---&
\makebox[3cm][c]{$\omega^2$}
\\
\hline
\end{tabular}
\end{center}
\end{table}
%
%
%
\begin{table}[thbp]
\caption{$\displaystyle G_{**~ *}
~/~\left(\frac{l\mu}{4e^2 b^2(i\omega-D_Hk^2)}\right)$. 
}
\label{tab:Gh-a_vector}
\begin{center}
\begin{tabular}{|c||c|c|}
\hline
&$tx$&$zx$
\\
\hline
\hline
$x$&
\makebox[3cm][c]{$-2i\omega$}
&\makebox[3cm][c]{$\displaystyle\frac{b}{(1+a)}\omega k$}
\\
\hline
\end{tabular}
\end{center}
\end{table}
%
%
%
\begin{table}[htbp]
\caption{$\displaystyle 
G_{*~ *}~/~\left(
\frac{l}{4e^2(1+a)b^2}\left(\frac{3a}{i\omega-D_Hk^2}
-\frac{(2-a)^2b}{2(1+a)}\right)\right)$. 
}
\label{tab:Ga-a_vector}
\begin{center}
\begin{tabular}{|c||c|}
\hline
&$x$
\\
\hline
\hline
$x$&\makebox[3cm][c]{$i\omega$}
\\
\hline
\end{tabular}
\end{center}
\end{table}
%
%
Notice that
$l^3/\kappa^2 = N_c^2/(4\pi^2)$ and 
 $l/e^2=N_c^2/(16\pi^2)$ for the R-charge,  
$l/e^2=N_cN_f/(4\pi^2)$ 
for the brane charge.
From the Green function for $U(1)$ currents $G_{x\ x}(\omega, k)$, 
one can read off the thermal conductivity $\kappa_T$ via Kubo 
formula, 
\begin{equation}
\kappa_T
\equiv
-\frac{(\epsilon+p)^2}{\rho^2T}\lim_{\omega\rightarrow 0}
\frac{1}{\omega}\mbox{Im}\Big(G_{x\ x}(\omega, k=0)\Big)
=2\pi^2\frac{N_c}{N_f}\frac{\eta T}{\mu^2}. 
\end{equation}
%

\subsection{Tensor type perturbation}

In the tensor type perturbation, an independent variable 
is just 
$$
h_{xy}(x). 
$$
By using Kubo formula, 
one can obtain the shear viscosity $\eta$ as 
\begin{equation}
\eta
\equiv-\lim_{\omega\rightarrow 0}\frac{1}{\omega}
\mbox{Im}\Big(G_{xy \ 
 xy}(\omega, k=0)\Big)
=\frac{l^3}{16\kappa^2b^3}, 
\end{equation}
where the retarded Green function is given by 
%
%
\begin{table}[htbp]
\caption{$\displaystyle 
G_{**~ **}~/~\left(
-\frac{l^3}{16\kappa^2b^3}\right)$. 
}
\begin{center}
\begin{tabular}{|c||c|}
\hline
&$xy$
\\
\hline
\hline
$xy$&\makebox[3cm][c]{$i\omega+bk^2$}
\\
\hline
\end{tabular}
\end{center}
\end{table}
%
%

\noindent
One can confirm the universal (within Einstein gravity\footnote{
If one consider higher derivative corrections to Einstein gravity,  
the viscosity bound could be 
modified~\cite{kp, blmsy1, nd, gmsst2, cns}. 
}
) ratio 
that is the ratio of the shear viscosity to the entropy density $s$, 
\begin{equation}
\frac{\eta}{s}=\frac{1}{4\pi}. 
\end{equation}

\section{Perturbative Solutions for $\widetilde{\Phi}_+$}

Substituting the equation (\ref{expansion_1}) into the 
equation (\ref{master_2}), 
one can read off one for $F_{+0}(u)$, 
\begin{equation}
0=
\Big(u^{-1}(1-u)(1+u-au^2)F_{+0}'\Big)'. 
\end{equation}
A general solution is given by 
\begin{equation}
F_{+0}(u)
=
C_0
+D_0
\Bigg\{
2\log\Big(1-u\Big)
-\log\Big(1+u-au^2\Big)
-\frac{3K_2(u)}{\sqrt{1+4a}}
\Bigg\}.
\end{equation}
Constants of integration $C_0$ and $D_0$ should be determined 
to be a regular function at the horizon. 
So we here choose $D_0=0$ and set 
\begin{equation}
F_{+0}(u)=C_0=C. \quad (\mbox{const.}) 
\end{equation}

By using this solution, one can get an equation for 
$F_{+1}(u)$ from the equation (\ref{master_2}), 
\begin{equation}
0=
\Big(u^{-1}(1-u)(1+u-au^2)F_{+1}'\Big)'
-i\frac{Cb(1+au^2)}{(2-a)u^2}. 
\end{equation}
A general solution is 
\begin{eqnarray}
F_{+1}(u)
&=&
C_1
+\frac{i
\Big(C(1-a)b
-(2-a) 
D_1\Big) 
\log\Big(1-u\Big)}
{(2-a)^2}
\nonumber\\
&&
+\frac{i 
\Big(Cb+(2-a)D_1
\Big) 
\log 
\Big(
1+u-au^2
\Big)}
{2(2-a)^2}
+\frac{3i 
\Big(
Cb+(2-a)D_1
\Big)
K_2(u)}
{2(2-a)^2 \sqrt{1+4a}}.
\quad  
\end{eqnarray}
Removing the singularity at the horizon, 
the constant $D_1$ should be 
$$
D_1=Cb\frac{1-a}{2-a}. 
$$
We also impose a boundary condition $F_{+1}(u=0)=0$ 
so as to fix the constant $C_1$.
Therefore the final form is 
\begin{equation}
F_{+1}(u)
=
\frac{iCb}{2(2-a)}
\left\{
\log\Big(1+u-a u^2\Big)
-\frac{6 K_1(u)}
{\sqrt{1+4a}}
\right\}.
\end{equation}

A differential equation for $G_{+1}(u)$ is 
\begin{equation}
0
=
\Big(
u^{-1}(1-u)(1+u-au^2)G_{+1}'
\Big)'
+\frac{Cb^2\Big(6-(1+a)u^2\Big)}{3(1+a)u^4}. 
\end{equation}
A general solution is 
\begin{eqnarray}
G_{+1}(u)
&=&
\tC_1
\nonumber 
\\
&&
+\frac{1}{6(2-a)(1+a)}
\Biggr\{-\frac{4C(2-a)b^2}{u}
\nonumber 
\\
&&
\hspace*{35mm}
+\frac{\left(C(1+5a-2a^2)b^2
-9(1+a)\widetilde{D}_1\right) K_2(u)}{
\sqrt{1+4a}}
\nonumber
\\
&&
\hspace*{35mm}
+\left(C(1-a)b^2+3(1+a)\widetilde{D}_1\right)
\nonumber 
\\
&&
\hspace*{50mm}
\times
\Bigg(
2\log\Big(1-u\Big)
-
\log\Big(1+u-au^2
\Big)
\Bigg)
\Biggr\}
\quad
\end{eqnarray}
and the constant $\widetilde{D}_1$ can be fixed as 
$$
\widetilde{D}_1
=-\frac{C(1-a)b^2}{3(1+a)}. 
$$ 
From the condition $\left.(uG_{+1})'\right|_{u=0}=0$, we can fix the constant
$\widetilde{C}_1$. 
The final form of the solution becomes 
\begin{equation}
G_{+1}(u)
=
\frac{2}{3}Cb^2
\left\{
\frac{K_1(u)}{\sqrt{1+4a}}-\frac{1}{(1+a)u}
\right\}. 
\end{equation}

A differential equation for $F_{+2}(u)$ is 
\begin{eqnarray}
0
&=&
\Biggl[
u^{-1}(1-u) \left(1+u-a u^2\right) 
\nonumber
\\
&&
\hspace*{5mm}
\times
\Bigg(\frac{Cb^2\log\Big(1-u\Big)}{(2-a)^2 (1-u)}
+\frac{ibF_{+1}(u)}{(2-a)(1-u)}
-\frac{ib \log (1-u) F_{+1}'(u)}{2-a}+F_{+2}'(u)\Bigg)
   \Biggr]'
\nonumber
\\
&&
-\frac{Cb^2}{u^2(1-u)(1+u-au^2)}. 
\end{eqnarray}
Integrating over $u$, we have 
\begin{eqnarray}
F_{+2}'(u)
&=&
\frac{1}{2(2-a)^2 \sqrt{1+4a} (1-u)(1+u-au^2)}
\nonumber 
\\
&&
\hspace*{10mm}
\times
\Biggl\{
2\sqrt{1+4a}(2-a)^2
\left(Cb^2-D_2 u\right)
\nonumber\\
&&
\hspace*{16mm}
-3Cb^2K_2(0)\left(1+u-au^2\right)
\nonumber 
\\
&&
\hspace*{16mm}
+Cb^2
\Big(
3+(5+3a-6a^2+2a^3)u-3au^2
\Big)
K_2(u)
\nonumber\\
&&
\hspace*{16mm}
+C\sqrt{1+4a}b^2
(1-u)(1+au) 
\log\Big(1+u-a u^2\Big)
\Biggr\}
\end{eqnarray}
and the constant $D_2$ can be fixed as 
$$
D_2
=
C b^2
\Big\{
1
-\frac{3K_2(0)}{2(2-a) 
\sqrt{1+4a}}
+\frac{(1+a)K_2(1)}{\sqrt{1+4a}}
\Big\}, 
$$ 
so that 
$\left.\left((1-u)F_{+2}'(u)\right)\right|_{u=1}=0$. 
Then, we find 
\begin{equation}
F'_{+2}(u)=Cb^2+\mathcal O(u). 
\end{equation}
Hence, we can write the solution of $F_{+2}(u)$ as 
\begin{eqnarray}
F_{+2}(u)
&=&
\!\int^u\!\dd u
\frac{Cb^2}{(1-u)(1+u-au^2)}
\nonumber 
\\
&&
\hspace*{10mm}
\times
\Bigg\{
1-u+\frac{(1-u)(1+au)\log\Big(1+u-a u^2\Big)}
{2(2-a)^2}
\nonumber 
\\
&&
\hspace*{16mm} 
-\frac{3K_2(0)(1-u)(1+au)}{2 (2-a)^2 \sqrt{1+4a}}
-\frac{(1+a)K_2(1)u}{\sqrt{1+4a}}
\nonumber 
\\
&&
\hspace*{16mm}
+\frac{\Big(3+(5+3a-6a^2+2a^3)u-3au^2\Big)K_2(u)}
{2(2-a)^2\sqrt{1+4a}}
\Biggr\}, 
\end{eqnarray}
which satisfies a boundary condition $F_{+2}(0)=0$. 

\section{Perturbative Solutions for $\widetilde{\Phi}_-$}

For $F_{-0}(u)$, one can get an equation 
\begin{equation}
0=\Bigg(u(1-u)(1+u-au^2)\Big(1+a-\frac{3}{2}au\Big)^{-2}F_{-0}'\Bigg)'.
\end{equation}
A general solution is given by 
\begin{eqnarray}
F_{-0}(u)
&=&
C_0
\nonumber 
\\
&&
+D_0
\Bigg\{
(2-a)\log\Big(1-u\Big)
-4(1+a)^2\log\big(u\big)
\nonumber 
\\
&&
\hspace*{11mm}
+\frac{1}{2}(2+a)(1+4a)\log\Big(1+u-au^2\Big)
-\frac{\sqrt{1+4a}(2+5a)K_2(u)}{2}
\Bigg\}. 
\nonumber 
\\
\end{eqnarray}
Since the function $F_{-0}(u)$ should be regular at the horizon, 
we choose $D_0=0$ and get 
\begin{equation}
F_{-0}(u)=C_0=\widetilde{C}. \quad (\mbox{const.})
\end{equation}

Substituting the solution to the equation (\ref{master_2}), 
we get an equation for $F_{-1}(u)$, 
\begin{eqnarray}
0
&=&
\Bigg(u(1-u)(1+u-au^2)\Big(1+a-\frac{3}{2}au\Big)^{-2}F_{-1}'\Bigg)'
\nonumber
\\
&&
+
\frac{i\widetilde{C}b\Big(2(1+a)+(4+7a)u-6a(1+a)u^2+3a^2u^3\Big)}
{2(2-a)\Big(1+a-\displaystyle\frac{3}{2}au\Big)^3}.
\end{eqnarray}
A general solution is given as 
\begin{eqnarray}
F_{-1}(u)
&=&
C_1
\nonumber 
\\
&&
-\frac{i}{54 (2-a)a^2 
}
\nonumber\\
&&
\hspace*{10mm}
\times
 \Biggl\{8
(1+a)^2 
\left(
\tC(1+4a)b
-27D_1(2-a)a^2
\right)\log (u)
\nonumber\\
&&
\hspace*{16mm}
-2
\left(\tC(2+7a+23a^2)b-27D_1(2-a)^2 a^2\right) 
\log\Big(1-u\Big)
\nonumber\\
&&
\hspace*{16mm}
-(2+a)(1+4a)
\left(
\tC(1+4a)b
-27D_1(2-a)a^2
\right) 
\log 
\Big(1+u-au^2
\Big)
\nonumber\\
&&
\hspace*{16mm}
+\sqrt{1+4a}(2+5a) 
\left(
\tC(1+4a)b
-27D_1(2-a)a^2
\right) K_2(u)
\Biggr\}. 
\end{eqnarray}
The constant of integration $D_1$ should be 
$$
D_1=\frac{\widetilde{C}b(2+7a+23a^2)}{27(2-a)^2a^2},  
$$
so that the singularity at the horizon could be removed.  
In addition, we require the condition 
$$
\left[F_{-1}(u)
-\log(u) \lim_{u\rightarrow0}\left(\frac{F_{-1}(u)}{\log(u)}\right)
\right]_{u=0}=0
$$ 
to fix the constant $C_1$. 
Then we get the final form of the solution
\begin{eqnarray}
F_{-1}(u)
&=&
\frac{i\tC b}{2(2-a)^2}
\Bigl\{ 
8(1+a)^2\log (u)
-(2+a)(1+4a)
\log\Big(
1+u-au^2
\Big)
\nonumber\\
&&
\hspace{20mm}
-2\sqrt{1+4a}(2+5a) K_1(u)
\Bigr\}.
\end{eqnarray}

Similarly we have a differential equation for $G_{-1}(u)$, 
\begin{eqnarray}
0
&=&
\Bigg(
u(1-u)(1+u-au^2)\Big(1+a-\frac{3}{2}au\Big)^{-2}
G_{-1}'
\Bigg)'
\nonumber 
\\
&&
+\frac{\widetilde{C}b^2}{8(1+a)\Big(\displaystyle 1+a-\frac{3}{2}au\Big)^5}
\Bigg\{
-4(1+a)(2+6a+3a^2+2a^3)
\nonumber 
\\
&&
\hspace*{51mm}
+2a(10+30a+57a^2+10a^3)u
\nonumber 
\\
&&
\hspace*{51mm}
-42a^2(1+a)^2u^2
+21a^3(1+a)u^3
\Bigg\}.
\end{eqnarray}
A general solution of this equation can be obtained 
\begin{eqnarray}
G_{-1}(u)
&=&
\tC_1
-\frac{\tC a^2b^2}{a(1+a)
\Big(\displaystyle 
1+a-\frac{3}{2}ua
\Big)}
\nonumber 
\\
&&
+\frac{1}{54a(1+a)(2-a)}
\nonumber 
\\
&&
\hspace*{10mm}
\times
\Bigg\{
4(1+a)^2(2-a)\left(7 \tC b^2+54\tD_1 a(1+a)\right) 
\log(u)
\nonumber
\\
&&
\hspace*{16mm}
-\left(2\tC(14+13a+17a^2)b^2
+54\tD_1(2-a)^2a(1+a)
\right)
\log\Big(1-u\Big)
\nonumber
\\
&&
\hspace*{16mm}
-(2+a)
\left(
\tC(7+11a-14a^2)b^2
+27\tD_1a(2-a)(1+a)(1+4a)
\right)
\nonumber 
\\
&&
\hspace*{35mm}
\times
\log
\Big(1+u-a u^2
\Big)
\nonumber
\\
&&
\hspace*{16mm}
+\frac{1}{\sqrt{1+4a}}
\Big(
\tC(14+57a+81a^2-16a^3)b^2
\nonumber 
\\
&&
\hspace*{37mm}
+27\tD_1a(2-a)(1+a)(1+4a)(2+5a) 
\Big)
K_2(u)
\Bigg\}. 
\end{eqnarray}
The constant of integration $\widetilde{D}_1$ might be fixed to remove
the singularity at $u=1$, 
$$
\tD_1
=
-\frac{\widetilde{C}b^2(14+13a+17a^2)}{27a(1+a)(2-a)^2}.
$$
Another constant of integration $\widetilde{C}_1$ is fixed to satisfy
the condition 
$$
\left[G_{-1}(u)
-\log(u) \lim_{u\rightarrow0}\left(\frac{G_{-1}(u)}{\log(u)}\right)
\right]_{u=0}=0.
$$
The final result of the solution is 
\begin{eqnarray}
G_{-1}(u)
&=&
\frac{1}{3}\tC b^2
\Biggl\{
-\frac{9a^2u}{2(1+a)^2\Big(\displaystyle 1+a-\frac{3}{2}au\Big)}
-\frac{6(1+a)(2+a)}{(2-a)^2}\log(u) 
\nonumber
\\
&&
\hspace{13mm}
+\frac{3(1+a)(2+a)}
{(2-a)^2}\log \Big(1+u-au^2\Big)
+\frac{6 (2+5a+6a^2)}{(2-a)^2\sqrt{1+4a}}
K_1(u)
\Biggr\}.\quad
\end{eqnarray}

A differential equation for $F_{-2}(u)$ is 
\begin{eqnarray}
0
&=&
\Biggl[
\frac{u(1+u-au^2)}
{(2-a)^2\Big(\displaystyle 1+a-\frac{3}{2}au\Big)^2}
\nonumber 
\\
&&
\hspace*{10mm}
\times
\Biggl(
(2-a)^2(1-u) F_{-2}'(u)
+i(2-a)b F_{-1}(u)
\nonumber 
\\
&&
\hspace*{17mm}
-i(2-a)b(1-u)\log\Big(1-u\Big)F_{-1}'(u)
+\tC b^2\log\Big(1-u\Big)
\Biggr)
\Biggr]'
\nonumber
\\
&&
+\frac{\tC b^2}
{\Big(\displaystyle 1+a-\frac{3}{2}au\Big)^2 (1-u) \left(1+u-au^2\right)}. 
\end{eqnarray}
Integrating over $u$, we have 
\begin{eqnarray}
F_{-2}'(u)
&=&
\frac{\Big(\displaystyle 1+a-\frac{3}{2}au\Big)^2}
{(1-u) u \left(1+u-a u^2\right)}
\nonumber 
\\
&&
\hspace*{10mm}
\times
 \Biggl\{D_2
 +\frac{18\tC ab^2}{(2-a)^2(1+4a)\Big(\displaystyle 1+a-\frac{3}{2}au\Big)}
\nonumber
\\
&&
\hspace{17mm}
+\frac{4\tC(1+a)^2b^2 u (1+u-au^2) \log (u) }
{(2-a)^3 \Big(\displaystyle 1+a-\frac{3}{2}au\Big)^2}
\nonumber
\\
&&
\hspace*{17mm}
-\frac{2\tC b^2}{(2-a)^3}
\left(
1+\frac{(2+a)(1+4a)u(1+u-au^2)}
{4\Big(\displaystyle 1+a-\frac{3}{2}au\Big)^2}
\right) 
\log
\Big(1+u-au^2\Big)
\nonumber
\\
&&
\hspace*{17mm}
-\frac{\tC\sqrt{1+4a}(2+5a)b^2K_2(0)u(1+u-au^2)}
{2(2-a)^3 \Big(\displaystyle 1+a-\frac{3}{2}au\Big)^2}
\nonumber
\\
&&
\hspace*{17mm}
+\frac{2\tC b^2}{(2-a)^3(1+4a)^{3/2}}
\Bigg(
1-2a(5+a)
\nonumber 
\\
&&
\hspace*{57mm}
+\displaystyle 
\frac{(2+5a)(1+4a)^2u(1+u-au^2)}
{4\Big(\displaystyle 1+a-\frac{3}{2}au\Big)^2}
\Bigg)
K_2(u)
\Biggr\}, 
\nonumber 
\\
\end{eqnarray}
and the constant $D_2$ can be fixed as 
\begin{eqnarray*}
D_2
&=&
\frac{4\tC b^2}{(2-a)^4(1+4a)^{3/2}}
\Biggl\{
\sqrt{1+4a}\Big(2(1+4a)(1+a)^2\log(2-a)-9a(2-a)\Big)
\nonumber
\\
&&
\hspace{36mm}
+\frac{1}{2}(2+5a)(1+4a)^2 K_2(0)
-(1+a)(2-2a+41a^2)K_2(1)
\Biggr\}, 
\end{eqnarray*}
so that $\left.\left((1-u)F_{-2}'(u)\right)\right|_{u=1}=0$. 
Then, we can write the solution as 
\begin{eqnarray}
F_{-2}(u)
&=&
\!\int^u\!\dd u
\frac{\tC b^2}{2(2-a)^4(1+4a)^{3/2}(1-u)u 
\left(1+u-au^2\right)}
\nonumber\\
&&
\hspace*{10mm}
\times
\Biggl\{8(2-a)(1+a)^2(1+4a)^{3/2}u(1+u-au^2)\log (u)
\nonumber
\\
&&
\hspace*{17mm}
+(2-a)(1+4a)^{3/2}
\Bigl(a(2+a)(1+4a)u^3
-(2+9a+13a^2)u^2
\nonumber
\\
&&
\hspace{54mm}
-(2-3a-8a^2)u
-4(1+a)^2
\Bigr) 
\log\Big(1+u-a u^2\Big)
\nonumber
\\
&&
\hspace*{17mm}
+(1+4a)^2(2+5a)K_2(0)(1-u) 
\Bigl(4(1+a)^2+(2-3a-8a^2)u
\nonumber 
\\
&&
\hspace*{77mm}
-a(2-a)u^2
\Bigr)
\nonumber
\\
&&
\hspace*{17mm}
-8(1+a)(2-2a+41a^2)K_2(1)\Big(1+a-\frac{3}{2}au\Big)^2
\nonumber
\\
&&
\hspace*{17mm}
-(2-a) 
\Bigl(a(1+4a)^2(2+5a)u^3
-(2-a)(1+11a+46a^2+18a^3)u^2
\nonumber
\\
&&
\hspace*{35mm}
-a(1+11a+46a^2+18a^3)u
\nonumber 
\\
&&
\hspace*{35mm}
-4(1+a)^2(1-10a-2a^2)
\Bigr)K_2(u)
\Biggr\}.
\end{eqnarray}
Expanding this expression around $u=0$, we have
\begin{equation}
F_{-2}(u)
=\tC b^2 D_-(1+a)\log(u)+C_2+\mathcal O(u),  
\end{equation}
where $C_2$ is an integration constant. 


\end{document}